%% file: MetabolicUberGraphs.tex
\documentclass{svmult}
\usepackage{amssymb,amsmath,amsfonts,graphics,graphicx,epsfig,multirow,color,subfig,stmaryrd,amsbsy,mathtools,xargs,xcolor,tikz}
\usetikzlibrary{shadows,shapes.geometric, decorations.markings,decorations.pathreplacing, calc, arrows}
                                                                                                                     
\newcommand\scalemath[2]{\scalebox{#1}{\mbox{\ensuremath{\displaystyle #2}}}}
\renewcommand{\thesection}{\arabic{section}}
\newcommand{\R} {\mathbb{R}}

\newcommand{\Ptt}[1]{\left( #1 \right)}

\newcommandx{\gw}[4][1=a,2=b,3=p]{W^{#1,#2}_{#3}\Ptt{#4}}

\newcommand{\bi}{\begin{itemize}}
\newcommand{\ei}{\end{itemize}}
\newcommand{\be}{\begin{enumerate}}
\newcommand{\ee}{\end{enumerate}}
\newcommand{\bd}{\begin{description}}
\newcommand{\ed}{\end{description}}

\newcommand{\bqn}{\begin{eqnarray}}
\newcommand{\eqn}{\end{eqnarray}}
\newcommand{\eqnn}{\nonumber\end{eqnarray}}
\newcommand{\eqnl}[1]{\label{#1}\end{eqnarray}}
\newcommand{\brem}{\begin{remark}}
\newcommand{\erem}{\end{remark}}
\newcommand{\bdeff}{\begin{definition}}
\newcommand{\edeff}{\end{definition}}
\newcommand{\ba}[1]{\begin{array}{#1}}
\newcommand{\ea}{\end{array}}

\newcommand{\bp}{\begin{proposition}}
\newcommand{\ep}{\end{proposition}}

\begin{document}

\title{Metabolic graphs, LIFE method and the modeling of drug action
on \textit{Mycobacterium tuberculosis}}

\titlerunning{Metabolic graphs model drug action on MTB using LIFE}

\author{Sean T. McQuade\inst{1}\and
Nathaniel J. Merrill\inst{1}\and
Benedetto Piccoli\inst{2}}

\institute{Center for Computational and Integrative Biology
Rutgers University--Camden, 303 Cooper St, Camden, NJ, USA,
\texttt{sean.mcquade@rutgers.edu}, \texttt{nathaniel.j.merrill@rutgers.edu}
\and
Department of Mathematical Sciences and
Center for Computational and Integrative Biology,
Rutgers University--Camden, 303 Cooper St, Camden, NJ, USA, \texttt{piccoli@camden.rutgers.edu}}

\maketitle

\abstract{This paper serves as a framework for designing advanced models for drug action on metabolism. Drug treatment may affect metabolism by either enhancing or inhibiting metabolic reactions comprising a metabolic network. We introduce the concept of
\textit{metabolic graphs}, a generalization of hypergraphs having specialized features common to metabolic networks. 
Linear-in-flux-expression (briefly LIFE) is a methodology for analyzing metabolic networks and simulating virtual patients. We extend LIFE dynamics to be compatible with metabolic graphs, including the more complex interactions of enhancer and inhibitor molecules that affect biochemical reactions. We discuss results considering network structure required for existence and uniqueness of equilibria on metabolic graphs and show simulations of drug action on \textit{Mycobacterium tuberculosis} (briefly MTB).}

\section{Introduction}\label{sec:intro}
The modeling of metabolic networks plays a crucial role in systems biology
and has many diverse applications, including in Quantitative Systems Pharmacology \cite{perez2015using}
for drug discovery and optimization of drug treatments.
There are various challenges at modeling, including the complexity and dimensionality of the involved networks, often times comprising hundreds of metabolites and thousands of reactions, enzymes and genes. 
For this reason, only methods corresponding to linear dynamics
presented the characteristic of scalability and computability to address such problems. 
In this area, Flux Balance Analysis (briefly FBA) plays a special role
for its simplicity and many successful uses, see 
\cite{gianchandani2010application,kauffman2003advances,lee2006flux,orth2010flux,raman2009flux}.
Other linear techniques proved efficient, such as zero deficiency theory,
Markov chains, Laplacian dynamics \cite{A06,bullo2016lectures,cinlar2013introduction,feinberg1974dynamics,gunawardena2012linear, jacquez1993qualitative}.\\
Beside the limitation of dealing only with linear dynamics, such methods
mostly neglected the nonlinear effects related to the action of enzymes,
the regulation effect of genes and the action of drugs on genes. 
In mathematical terms, even if one accepts a linear dynamics in terms of the metabolites, the action of enzymes, genes and drugs affects fluxes among metabolites.  For the example of a downregulation action of a gene, then
reducing a flux would necessarily correspond to a nonlinear term in
the dynamics of the involved metabolites. Even more, such action
cannot be represented in the usual language of graph theory,
thus requiring the use of more general theories.\\
The present paper addresses such limitations using two main tools:\\
1. A new representation of metabolite dynamics called 
\emph{Linear-In-Flux-Expression} (briefly LIFE) to allow nonlinearities in the dynamics.\\
2. The use of \emph{metabolic graphs}, based on hyper and uber-graphs, to allow the representation of enzymes, genes and drug action.\\
The rest of this Introduction details the two main methods and provides a brief description of our main application: the action
of antibiotics on the \textit{Mycobacterium tuberculosis} (briefly MTB).

The paper is organized as follows. Section \ref{sec:LIFE} discuss the LIFE approach and provide basic definition,
while in Section \ref{sec:CCM} we illustrate the central carbon metabolism network of MTB.
In particular the example shows the necessity of introducing the concept of metabolic graph which
is done in Section \ref{sec:metabolic_graphs}. Extension of LIFE to metabolic graphs is done in Section \ref{sec:EI}
and the problem of existence and uniqueness of equilibria is explored in Section \ref{sec:EQ}.
Application of the methods to drug action on MTB (specifically synthesis of antibiotics) is illustrated
in Section \ref{sec:MTB}.

\subsection{The LIFE method}
Recently, flux balance analysis techniques were expanded to include nonlinear metabolite dynamics. More precisely the usual way to write the dynamics
of a metabolic network is:
\begin{equation}
\dot x=S(f)\cdot x,
\end{equation}
where $x\in\R^n$ is the vector of metabolite levels, $f\in\R^m$ the vector of fluxes
and $S$ is a $n\times n$ matrix, called stoichiometric matrix, which link
the flux levels to the metabolite dynamics. The system is linear in the metabolites
and usually the system is represented by a directed graph $G=(V,E)$,
with $V$ the set of vertices (or nodes) representing metabolites and $E\subset V\times V$
the set of edges, representing fluxes linked to biochemical reactions.\\
Often times the metabolite dynamics may be nonlinear, e.g. given
by a Michaelis-Menten-type kinetics, however the linearity in the fluxes
is essentially given by definition. Moreover, many important biological
and mathematical problems can be formulated in terms of the fluxes levels,
which usually are assumed to be constant or evolve by slow dynamics
or, finally, modified by drug intervention. Thus the main idea behind
the Linear-In-Flux-Expression approaches (briefly LIFE) is to write the system as:
\begin{equation}
\dot x= S(x)\cdot f,
\end{equation}
where $x$ and $f$ are as before, but the new stoichiometric matrix
$S=\{S_{ve}\}_{v\in V,e\in E}$ is now an $n\times m$ matrix explicitly
encoding the linear behavior in the fluxes. A typical advantage is the possibility
of consider nonlinear dynamics allowing $S_{ve}$ to be nonlinear in $x$
and exploring the space of equilibria as function of fluxes, given by the kernel of the matrix $S$. 
General conditions on network topology (connection of every node to excretion, see \cite{merrill2019stability})
guarantee existence and uniqueness of an equilibrium $\bar{x}_f$ for every flux vector $f=\{f_e\}$,
thus network asymptotic dynamics is captured by the map $f\to \bar{x}_f$.
We will provide more detail below about the many results
achievable by this approach combining  several different methods for modeling chemical systems, including systems biology, zero deficiency theory, laplacian dynamics, and Markov chains \cite{palsson2015systems,A06,bullo2016lectures,jacquez1993qualitative,feinberg1974dynamics,gunawardena2012linear,cinlar2013introduction}. 
We will also refer the reader to
\cite{AMMP19,mcquade2017linear,merrill2019stability} 
for a general presentation of the LIFE approach.

\subsection{Hyper, uber and metabolic graphs}
As explained above FBA and other methods rely on representing
the metabolic network as a directed graph, where edges represent biochemical reaction. There are (at least) three main limitations related to representing
a complex metabolic network with a standard directed graph and these are:\\
1. Most networks include inflows and outflows (also called intakes and excretions in LIFE methodology) to the external environment or to other networks. 
Virtual nodes to represent such flow must be included or, alternatively,
one must include directed edges with a node only on one end.\\
2. Some biochemical reactions necessarily involve more than two
metabolites, e.g. when two or more compounds interact to form
a set of other compounds. Therefore edges with multiple entering
and exiting nodes must be included.\\
3. The action of enzymes, genes and drugs often times affect
a specific reaction acting as enhancer or inhibitors. Such actions
can be represented by edges joining a node to another edge.\\
\begin{figure}[h!]
\hspace{2.3cm}\input{intro.tex}
\caption{\footnotesize We define a \textit{metabolic graph} to have two added features compared to simple directed graphs: 1. \textbf{Left} weighted hyperedge $h$ will replace simple edges. These edges have weights assigned to each branch of the hyperedge respecting the stoichiometry of the corresponding reaction; and 2. \textbf{Right} enhancer or inhibitor dynamics acting on edge $e$. The inhibitor (enhancers) are included to model the action of molecules inhibiting (promoting) the enzyme for a reaction corresponding to edge $e$. The edges $u_1,u_2$ are called uberedges and connect a node to an edge. }\label{fig:summary_figure} 
\end{figure}
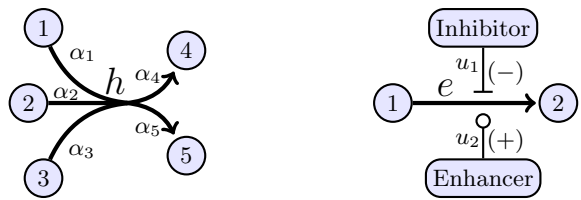
All these extensions can be achieved by introducing appropriate
generalization of the concept of graph. In particular hypergraphs \cite{bretto2013hypergraph,voloshin2009introduction}
contain hyperedges with multiple nodes, ubergraphs \cite{joslyn2017ubergraphs} include
uberedges connecting node to other edges. We use this tool
to defined a generalized graph, called metabolic graph addressing
the limitations 1-3.
Figure \ref{fig:summary_figure} depicts the main idea behind the definition
of metabolic graph.
Metabolic graphs are hypergraphs which may include an uberedge which connects a node to a hyperedge. 

\subsection{Tuberculosis}
The \textit{Mycobacterium tuberculosis} (briefly MTB) has infected thirty percent of the world's population according to the World Health Organization(WHO). The WHO declared tuberculosis (brifely TB) a global emergency in 1993 \cite{nakajima1993tuberculosis}.
The bacterium is known to endure hostile environments within the host organism through two main factors 1. genetically diverse sub-populations \cite{cilfone2013multi,balazsi2008temporal} and 2. a sophisticated gene regulatory network(briefly GRN) \cite{balazsi2008temporal,galagan2013mycobacterium,peterson2014high}.
Developing new TB drugs with improved effectiveness depends on modelling MTB in multiple metabolic states.


In the past, designing treatment for TB has been exceptionally difficult, often requiring drug cocktails of up to four different drugs. In addition to multiple drugs, treatment may last up to six months, and this duration along with complications due to drug side effects can negatively effect patient compliance. Two main approaches have been taken to improve TB treatment. 
the first approach is discovering more effective dosing regimens of these drug cocktails \cite{cokol2017efficient,mcilleron2006determinants,tostmann2013pharmacokinetics}

The second approach is to anticipate the response of the GRN to treatment\cite{stewart2001overexpression,kendall2007highly,delahaye2018alveolar}. This allows one to exploit the changing metabolism of MTB, but requires accurate models of the response to drug treatment.  Such models of MTB metabolism are difficult to construct due to the numerous complex effects of the GRN on the metabolic state as a consequence of environmental or internal chemical conditions. 

In either approach, it is essential to model the individual drug interactions carefully so that a model can be built with a solid foundation. Modeling drug interactions in an intuitive way requires us to allow structures where molecules or metabolites inhibit or enhance a biochemical reaction. This action can be depicted as a node attached to a hyperedge (connecting nodes which represent multiple reactant metabolites to nodes representing products of the reaction).

\section{Modeling metabolic networks with LIFE}\label{sec:LIFE}
Generalizing the dynamics of FBA, we focus on the following class of systems:
\begin{equation}\label{eq:first-dyn-LIFE} 
\dot{x} = S(x)\cdot f,
\end{equation}
where the stoichiometric matrix $S$ is as in FBA models, but depending on $x$,
and $f$ is the column vector of fluxes.  
The values $x_1,x_2 \ldots$ represent mass of metabolties and so are constrained to $\mathbb{R}_+$.
The network is associated to a directed graph $G=(V,E)$, with $V=\{v_1,\ldots,v_n\}$ the set of nodes
and $E\subset V\times V$ the set of directed edges.
$S_{v e}(x)$ indicates the matrix entry corresponding to node $v$ and edge $e$.  
The network equilibria correspond to the null space of $S(x)$ thus depend both on the fluxes and metabolite levels
and a detailed analysis is provided in \cite{merrill2019stability} combining various mathematical approaches as follows.
Linear systems without intakes nor excretions are related to \emph{continuous-time Markov chains} \cite{cinlar2013introduction},
while linear systems with intakes and excretions are known as \emph{compartmental systems} \cite{bullo2016lectures,jacquez1993qualitative}. Nonlinear systems could be treated by using the results of \cite{maeda1978asymptotic}. Metabolic networks contain exchange fluxes representing incoming mass from other parts of the network, or from the outside environment, and they are necessary for the existence of equilibria. For traditional metabolic networks described by a directed graph $G=(V,E)$, directed edges from a virtual node $v_0$ which acts as a source to some node in the graph are called intakes, and edges from a node in the graph to some virtual node $v_{n+1}$ acting as a sink are called excretions.
\begin{definition}
Given a directed graph $G=(V,E)$, an edge $(v_0,v_1)\in E$ 
for the virtual node $v_0$, the edge $(v_0, v_1)$ is called an intake, and $v_1$ is an intake node.  For the virtual node $v_{n+1}$ an edge $(v_j,v_{n+1})$ is called an excretion, and $v_j$ is an excretion node. The set of intake nodes is denoted $I$, and the set of excretion nodes is denoted by $J$.
\end{definition}
A general assumption, still allowing significant results, is the following:
\begin{equation*}\label{eq:H1}
(H1) \hspace{2cm}
S_{v e}(x) = \begin{cases}
 -F_e(x_v)  & e=(v,w),\ v \in V, w \in V\cup \{v_{n+1}\}\\
 F_e(x_w) & e=(w,v),\ w \in V \\
 1 & e=(v_0,v)\ v\in I \\
0 & \mbox{otherwise},
\end{cases}\hspace{3cm}
\end{equation*}
where $F_e: \mathbb{R}_+\to \mathbb{R}_+$ is differentiable, strictly increasing, with $F_e(0)=0$, for $\mathbb{R}_+ = \{x\in \mathbb{R}: x \geq 0\}$.  (H1) is a natural assumption on the system; the flow from a metabolite will depend only on that metabolite, but for a metabolite with multiple edges may have different kinetics with each reaction.  This assumption also includes nonlinear kinetics such as Michaelis-Menten \cite{kou2005single} corresponding to
Hill functions $F_v(x_v)= \frac{x_v^{p}}{K + x_v^{p}}$ with $p \in \mathbb{N}$. Moreover, all columns of $S$ have zero sum, except those corresponding to intakes and excretions, which have positive and negative sum, respectively.
Therefore, the dynamics \eqref{eq:first-dyn-LIFE} can be interpreted as mass conservation law. To be biologically meaningful, we further
restrict to equilibria for which all components of $f$ and $x$
are positive, thus rendering the problem nonlinear even for linear dynamics. A previous result shows that under (H1), the existence of equilibria depends on the structure of the network.

\begin{proposition}\label{prop:A_neccessary}
Consider a system \eqref{eq:first-dyn-LIFE} satisfying (H1).
Assume there exists an equilibrium $\bar{x}\in (\R_+)^n$ 
for a flux vector $f$ such that $f_e>0$ for every $e\in\tilde  E$.
Then for every node $v \in V$ for which there exists a path from $I$ to $v$, 
there exists a path from $v$ to $J$.
\end{proposition}
This proposition is proven in \cite{merrill2019stability}. We extend these results to metabolic networks including enhancers and inhibitors as explained in Section
\ref{sec:metabolic_graphs}.

\subsection{Central Carbon Metabolism in \textit{Mycobacterium tuberculosis}}\label{sec:CCM}
The dynamics described in the previous section can be used to model the flows of a metabolic network.
The entries of the stoichiometric matrix describe the mass flowing along an edge of the network per time.  
The flux for each edge represents the speed of this flow.
We present an example metabolic network from \cite{shi2010carbon} to motivate expanding the LIFE method from section \ref{sec:LIFE}.

The Metabolic network shown in Figure \ref{fig:central_carbon} is the central carbon metabolism of MTB derived from an infection in mouse. The network shown is a metabolic network of MTB in a state of growing bacilli. This network depicts carbon from lipid and sugar catabolism used by MTB for generating energy and biosynthetic precurors required for growth. Specifically glucose is a product of catabolism and the cell uses this energy to synthesize enzymes for the pentose phosphate pathway and provide ribose 5-P for nuclotide synthesis.  Glycolysis yields metabolites phosphoenolpyruvate(PEP), Pyruvate, and acetyl-CoA. Experiments showed that MTB preferentially uses fatty acids as a carbon source. When MTB is in the ``non-growing'' state within mouse infection, some of the genes required for this process are known to be upregulated. These observations led to the proposal that MTB switches its carbon source from sugars to fatty acids during the persistent phase of infection.


\tikzset{%
   peer/.style={draw,rectangle, rounded corners=.2cm,fill=blue!20}}
\tikzset{
  jump/.style={
     to path={
         let \p1=(\tikztostart),\p2=(\tikztotarget),\n1={atan2(\y2-\y1,\x2-\x1)} in
         (\tikztostart) -- ($($(\tikztostart)!#1!(\tikztotarget)$)!0.15cm!(\tikztostart)$)
         arc[start angle=\n1+180,end angle=\n1,radius=0.15cm] -- (\tikztotarget)}
  },
  jump/.default={0.5}
}

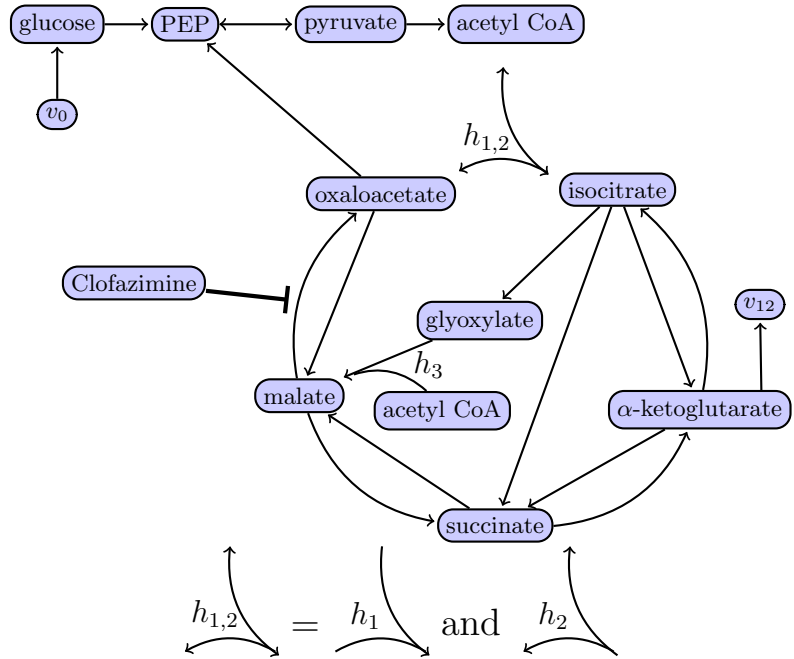
\begin{figure}[h]
\begin{center}
\def \n {5}
\begin{tikzpicture}[->,shorten >=1pt,auto,node distance=2.2cm, thick]
\node[peer]         (ketoglutarate)   at ({360/\n * (1 - 1)-20}:2.8){$\alpha$-ketoglutarate};
\node[peer]         (isocitrate)   at ({360/\n * (2 - 1)-20}:2.5){isocitrate};
\node[peer]         (oxaloacetate)   at ({360/\n * (3 - 1)-14}:2.5){oxaloacetate};
\node[peer]         (malate)   at ({360/\n * (4 - 1)-20}:2.8){malate};
\node[peer]         (succinate)   at ({360/\n * (5 - 1)-20}:2.5){succinate};
\node[peer,above of=isocitrate,xshift=-1.35cm] (acetyl_coa) {acetyl CoA};
\node[peer,left of=acetyl_coa,xshift = 0cm] (pyruvate) {pyruvate};
\node[peer,left of=pyruvate] (PEP) {PEP};
\node[peer,left of=PEP,xshift=0.5cm] (glucose) {glucose};
\node[peer, right of=malate,xshift=0.2cm,yshift=1cm] (glyoxylate) {glyoxylate};
\node[peer,below of=glyoxylate,xshift=-0.5cm,yshift=1cm] (acetyl_coa2) {acetyl CoA};
\node[peer,below of=glucose,yshift=1cm] (intake) {$v_0$};
\node[peer,above of=ketoglutarate,xshift=0.8cm,yshift=-0.8cm] (excretion) {$v_{12}$};
\node[peer,left of=malate,xshift=0cm,yshift=1.5cm] (Clofazimine) {Clofazimine};


    \path[-]
    (0.02,3.6) edge[<->,bend right, thick] node [xshift = 0.2cm,yshift = -0.3cm] {} (0.65,2.15);
    \path[-]
(0.6,2.2) edge[->,bend right, thick] node [xshift = -0.2cm,yshift = 0.6cm] {\large $h_{1,2}$} (-0.6,2.2);
\path[-]
    (PEP) edge[<->,thick] node[below,xshift = 0.2cm,yshift = 0cm] {} (pyruvate);
\path[-]
    (glucose) edge[->,thick] node[below,xshift = 0.2cm,yshift = 0cm] {} (PEP);
    \path[-]
    (pyruvate) edge[->,thick] node[below,xshift = 0.2cm,yshift = 0cm] {} (acetyl_coa);
        \path[-]
(malate) edge[->,thick,bend left] node[below,xshift = 0.2cm,yshift = 0cm] {} (oxaloacetate);
\path[-]
(oxaloacetate) edge[->,thick] node[below,xshift = 0.2cm,yshift = 0cm] {} (PEP);
\path[-]
    (isocitrate) edge[->,thick] node[below,xshift = 0.2cm,yshift = 0cm] {} (glyoxylate);
\path[-]
 (glyoxylate) edge[->,thick] node[below,xshift = 0.2cm,yshift = 0cm] {} (malate);
        \path[-]
(oxaloacetate) edge[->,thick] node[below,xshift = 0.2cm,yshift = 0cm] {} (malate);
    \path[-]
(succinate) edge[->,thick] node[below,xshift = 0.2cm,yshift = 0cm] {} (malate);
    \path[-]
(malate) edge[->,thick,bend right] node[below,xshift = 0.2cm,yshift = 0cm] {} (succinate);
    \path[-]
(succinate) edge[->,thick,bend right] node[below,xshift = 0.2cm,yshift = 0cm] {} (ketoglutarate);
    \path[-]
(ketoglutarate) edge[->, thick] node[below,xshift = 0.2cm,yshift = 0cm] {} (succinate);
    \path[-]
(isocitrate) edge[->, thick] node[below,xshift = 0.2cm,yshift = 0cm] {} (ketoglutarate);
    \path[-]
(isocitrate) edge[->, thick] node[below,xshift = 0.2cm,yshift = 0cm] {} (succinate);
    \path[-]
(ketoglutarate) edge[->, thick,bend right] node[below,xshift = 0.2cm,yshift = 0cm] {} (isocitrate);
%
    \path[-]
    (intake) edge[->, thick] node[below,xshift = 0.2cm,yshift = 0cm] {} (glucose);
     \path[-]
(3.45,-0.7) edge[->, thick] node[below,xshift = 0.2cm,yshift = 0cm] {} (excretion);
\path[-]
(acetyl_coa2) edge[ thick,bend right] node[above,xshift = 0.5cm,yshift = -0.2cm] {\large $h_3$} (-2,-.50);
\path[-]
(Clofazimine) edge[-|,ultra thick] (-2.8,.5) ;
\end{tikzpicture}

\begin{tikzpicture}
    \path[-]
    (-2,3.6) edge[<->,bend right, thick] node [xshift = 0.2cm,yshift = -0.3cm] {} (-1.35,2.15);
    \path[-]
(-1.4,2.2) edge[->,bend right, thick] node [xshift = -0.2cm,yshift = 0.3cm] {\large $h_{1,2}$} (-2.6,2.2);
\node[] at (-1,2.5) {\Large =};
    \path[-]
    (0.02,3.6) edge[->,bend right, thick] node [xshift = 0.2cm,yshift = -0.3cm] {} (0.65,2.15);
    \path[-]
(0.6,2.2) edge[-,bend right, thick] node [xshift = -0.2cm,yshift = 0.3cm] {\large $h_{1}$} (-0.6,2.2);
\node[] at (1.2,2.6) {\Large and };
\path[-]
(2.52,3.6) edge[<-,bend right, thick] node [xshift = 0.2cm,yshift = -0.3cm] {} (3.15,2.15);
    \path[-]
(3.1,2.2) edge[->,bend right, thick] node [xshift = -0.2cm,yshift = 0.3cm] {\large $h_{2}$} (1.9,2.2);
\end{tikzpicture}
\end{center}
\caption{The central carbon metabolism of \textit{Mycobacterium tuberculosis} with TB drug action shown. The well known drug Clofazimine inhibits the reaction which oxidizes malate. The edges labeled $h_{1,2}$ represents two different hyperedges, and $h_3$ is a third hyperedge connecting three metabolites.}\label{fig:central_carbon}
\end{figure}

\vspace{-0.5cm}
\noindent
Simple directed edges, as in standard graphs, model most reactions in carbon metabolism. However,
the complete dynamics involves several hyperedges, i.e. generalized edges connecting more than two nodes.
The hyperedge labeled $h_{1,2}$ consists of two distinct directed hyperedges,
each connecting three metabolites involved in a single chemical reaction 
(the first having two reactants and the second having two products). 
The details about the reactions are taken from the KEGG database \cite{kanehisa2000kegg}. 
In particular, $h_1$ describes a chemical reaction producing isocitrate from Oxaloacetate, and acetyl CoA, and $h_2$ uses isocitrate to produce acetyl CoA and oxaloacetate.
The third hyperedge $h_3$ models acetyl CoA reacting with glyoxylate to produce malate.
We also included the drug Clofazimine, which acts as an inhibitor on the malate-oxaloacetate reaction.\\
The need to describe reactions depicted by the hyperedges and inhibitor in Figure \ref{fig:central_carbon} is what brings us to metabolic graphs and extending the LIFE method accordingly.

\section{Metabolic graphs}\label{sec:metabolic_graphs}
\input{EI.tex}

\section{Modeling Tuberculosis with metabolic graphs}\label{sec:MTB}
\begin{figure}[h]
\input{antibiotics_subgraph}
\caption{A subnetwork of the MTB synthesis of antibiotics. The drugs written show how they affect reactions (+ enhances the enzyme, - inhibits)}\label{fig:antibiotic_pathway}
\end{figure}
\begin{figure}[h]
\begin{center}
\includegraphics[scale=.33]{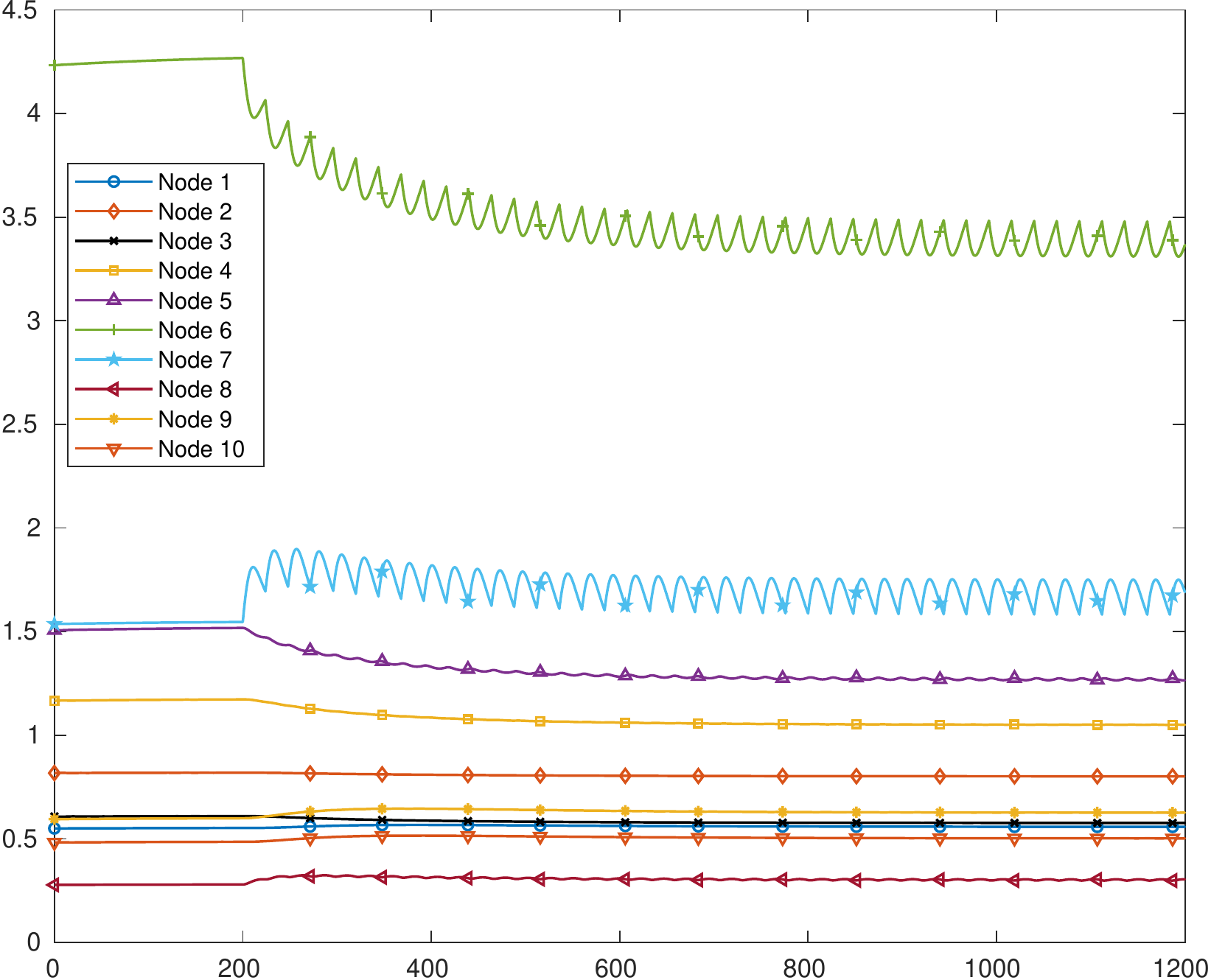}
\hspace{.5cm}
\includegraphics[scale=.33]{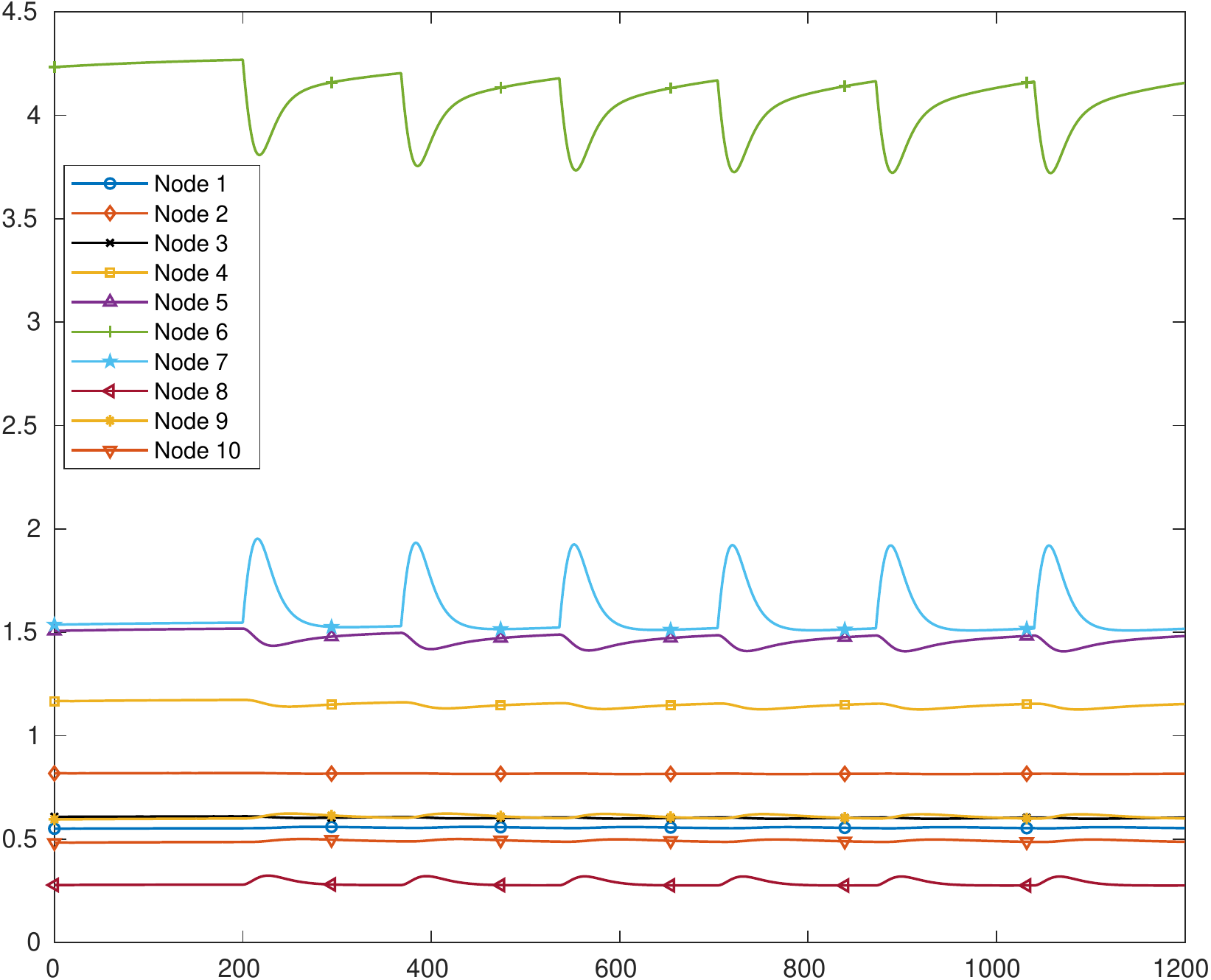}
\caption{\footnotesize{Evolution of MTB biosynthesis
of antibiotics.  The first 200 hours are without drug,
then results with (left) daily one hour treatment with Rifapentine and (right) weekly treatment at a higher dose.}}\label{fig:sim-stable}
\end{center}
\end{figure}

Here we focus on the synthesis of antibiotic network of MTB. Such a network is
shown in Figure \ref{fig:antibiotic_pathway} including the action of a set of antibiotics commonly used in cure of tuberculosis.
The action of drugs on edges is solely due to regulation of the enzyme for the corresponding reaction. We see that a group of three drugs regulates an enzyme for a reaction from node $v_1$ to node $v_{10}$, $f_{(v_{1}, v_{10)}}$. In this case treatment with any single drug (Amikacin, Capreomycin, or Clofazimine) will effect a single gene called ``Rv1475c.'' Amikacin and Capreomycin will upregulate the gene, and Clofazimine will down regulate the gene.  For edge $f_{(v_3,v_2)}$, Clofazimine downregulates two genes, both acting as enzymes for the corresponding reaction. For the excretion from node $v_2$ out of the subnetwork, four drugs (Amikacin, Clofazimine, Ethionamide, and Isoniasid) all upregulate the gene Rv0211. The combination of downregulating $f_{(v_3,v_2)}$, and upregulating the excretion from $v_2$ will serve to lower the metabolite represented by $v_2$, namely Oxalacetic acid (chemical formula: C4H4O5). 
Figure \ref{fig:sim-stable} compares two regimens with Rifapentine,
showing more effectiveness of thee daily versus the weekly treatment
on the biosynthesis of antibiotic network.\\
The dynamics of the biosynthesis of antibiotics network (Fig. \ref{fig:antibiotic_pathway}) can be written as \eqref{eq:first-dyn-LIFE}, where $S: \mathbb{R} \to M_{10\times 28}$ is a sparse matrix. 
\begin{equation}
\scalemath{.56}{
\left(\begin{array}{cccccccccccccccccccccccccccc} -x_{1} & x_{10} & 0 & 0 & 0 & 0 & 0 & 0 & 0 & 0 & 0 & 0 & 0 & 0 & 0 & 0 & 0 & 0 & 0 & 0 & -x_{1} & C_{1} & -x_{1} & x_{2} & 0 & 0 & 0 & 0\\ 0 & 0 & x_{3} & -x_{2} & -x_{2} & C_{2} & 0 & 0 & 0 & 0 & 0 & 0 & 0 & 0 & 0 & 0 & 0 & 0 & 0 & 0 & 0 & 0 & x_{1} & -x_{2} & 0 & 0 & 0 & 0\\ 0 & 0 & -x_{3} & x_{2} & 0 & 0 & 0 & 0 & 0 & 0 & -x_{3} & x_{4} & 0 & 0 & 0 & 0 & 0 & 0 & 0 & 0 & 0 & 0 & 0 & 0 & 0 & 0 & 0 & 0\\ 0 & 0 & 0 & 0 & 0 & 0 & 0 & 0 & 0 & 0 & x_{3} & -x_{4} & 0 & 0 & 0 & 0 & x_{5} & -x_{4} & 0 & 0 & 0 & 0 & 0 & 0 & 0 & 0 & C_{3} & -x_{4}\\ 0 & 0 & 0 & 0 & 0 & 0 & 0 & 0 & 0 & 0 & 0 & 0 & 0 & 0 & -x_{5} & x_{6} & -x_{5} & x_{4} & 0 & 0 & 0 & 0 & 0 & 0 & 0 & 0 & 0 & 0\\ 0 & 0 & 0 & 0 & 0 & 0 & 0 & 0 & 0 & 0 & 0 & 0 & x_{7} & -x_{6} & x_{5} & -x_{6} & 0 & 0 & 0 & 0 & 0 & 0 & 0 & 0 & 0 & 0 & 0 & 0\\ 0 & 0 & 0 & 0 & 0 & 0 & x_{8} & -x_{7} & 0 & 0 & 0 & 0 & -x_{7} & x_{6} & 0 & 0 & 0 & 0 & 0 & 0 & 0 & 0 & 0 & 0 & C_{4} & -x_{7} & 0 & 0\\ 0 & 0 & 0 & 0 & 0 & 0 & -x_{8} & x_{7} & x_{9} & -x_{8} & 0 & 0 & 0 & 0 & 0 & 0 & 0 & 0 & 0 & 0 & 0 & 0 & 0 & 0 & 0 & 0 & 0 & 0\\ 0 & 0 & 0 & 0 & 0 & 0 & 0 & 0 & -x_{9} & x_{8} & 0 & 0 & 0 & 0 & 0 & 0 & 0 & 0 & -x_{9} & x_{10} & 0 & 0 & 0 & 0 & 0 & 0 & 0 & 0\\ x_{1} & -x_{10} & 0 & 0 & 0 & 0 & 0 & 0 & 0 & 0 & 0 & 0 & 0 & 0 & 0 & 0 & 0 & 0 & x_{9} & -x_{10} & 0 & 0 & 0 & 0 & 0 & 0 & 0 & 0 \end{array}\right)}
\end{equation}

We show a simulation of the MTB biosynthesis of antibiotics network (Figure \ref{fig:antibiotic_pathway}) to illustrate what can happen. Figure \ref{fig:sim-stable} will converge to different periodic solutions based on the dosage chosen. The left hand side of Figure \ref{fig:sim-stable} shows the evolution with an hourly treatment of Rifapentine, and the right hand side shows the evolution with a weekly treatment. Note that most metabolites in the system converge to the same equilibrium value, however the metabolite labeled ``node 6'' has oscillations that are centered around a very different value depending on dosage.

%
%

\vfill\eject
\appendix
\section{Appendix: Definitions for general metabolic ubergraphs}
For completeness, this section constructs a more general version of a directed hypergraph, weighted directed hypergraph, and metabolic graph by building upon the classical notions of a graph and hypergraph. For the following definitions, $V$ is a set of nodes, and $\mathcal{P}({\cal S})$ is the power set of a set ${\cal S}$.

\begin{definition}
A \textit{directed graph} is a pair $G = (V,E)$ where $E$ is the set of ordered pairs such as $(v_i,v_j)$ with $v_i,v_j\in V$ which indicates a directed edge, starting from the initial node $v_i$ and pointing to the terminal node $v_j$.
\end{definition}

\begin{definition}
A weighted directed hyperedge is a couple ${\cal H}\ni h=(X,Y)$ with \
$X\in\mathcal{P}(V)$ or $X = \{v_0\}$, $Y\in\mathcal{P}(V\cup \{v_{n+1}\})$, and corresponding weights $\Psi_h: h \mapsto (\Psi_h^{\text{out}},\Psi_h^{\text{in}})$ where
$\Psi_h^{\text{out}}:X\mapsto \mathbb{R}_+$ and $\Psi_h^{\text{in}}:Y\mapsto \mathbb{R}_+$. 
\end{definition}

\begin{definition}
A weighted directed hypergraph $G$ is an ordered triplet $G=(V,{\cal H},\Psi_{\cal H})$ where $\Psi_{\cal H}$ denotes the set of functions assigning weights to hyperedges, i.e. $\Psi_{\cal H}=\{\Psi_h: h\in {\cal H}\}$.    
\end{definition}
As shown in Section \ref{sec:metabolic_graphs}, weighted directed
hypergraphs are the right mathematical object to describe
biochemical reactions involving multiple compounds. Interestingly,
while the theory of (not directed) hypergraphs seems well developed \cite{bretto2013hypergraph,voloshin2009introduction},
directed hypergraphs have been less explored in the literature.

We are now ready to introduce a general definition of ubergraphs.
The concept was introduced in \cite{joslyn2017ubergraphs}, but
the used definition appears to allow some pathological examples, thus
we provide an alternative definition below.\\
First, given a finite set of nodes $V=\{v_1,\ldots,v_n\}$ define recursively $P_k$ as: 
\begin{equation}
P_k =  \mathcal{P}\left(\bigcup_{i=0}^{k-1}P_i\right)\setminus \{ \emptyset \}, \quad P_0=V
\end{equation}
\begin{definition} A depth-$k$ ubergraph $U$ is a $k+1$-tuple $G = (U_0=V,U_1,\ldots,U_k)$ where $U_i\subseteq P_i$ is a finite set of uberedges
and $U_i\subset \mathcal{P}\left(\bigcup_{i=0}^{k-1}U_i\right)\setminus \{ \emptyset \}$. We call an element of $U_i$ a depth-$i$ uberedge. 
\end{definition}
Let us provide some details on such definition. First notice that
a depth-1 Ubergraph is a classical hypergraph. Indeed in this case
$P_1=\mathcal{P}\left(V\right)\setminus \{ \emptyset \}$ which
is precisely the definition of the set of hyperedges.\\
In \cite{joslyn2017ubergraphs} the definition was slightly different 
allowing $\emptyset$ as an hyperedge.
However, this give rise to the problem of the meaning in models
of an empty heperedge. The problem becomes more dramatic for
general depth $k$ ubergraphs. Another difference is that we force
$U_i$ to be subset of $\mathcal{P}\left(\bigcup_{i=0}^{k-1}U_i\right)\setminus \{ \emptyset \}$. If this is not the case, then we may have depth-$k$ uberedges
which connect uberedges of lower depths, without them being included
in the ubergraph, see the following discussion for depth-$2$ ubergraphs.\\
Depth-$2$ ubergraphs contains more general uberedges than hypergraphs.
More precisely, $G=(V,U_1,U_2)$, where $U_1$ is a set of hyperedges
and $U_2\subset \mathcal{P}\left(V\cap U_1\right)\setminus \{ \emptyset \}$.
Therefore $u\in U_2$ is a set $\{v_{k_1},\ldots,v_{k_m},h_{i_1,},\ldots,h_{i_p}\}$
where $v_{k_l}\in V$ and $h_{i_l}\in U_1$. A particular case
would be an uberedge $\{v,h\}$, $v\in V$, $h\in U_2$, which could model
an enhancer or inhibitor action of metabolite $x_v$ on the reaction
represented by the hyperedge $h$. Let us remark again
the importance of restricting the condition $U_i\subset P_i$ to the more
stringent we use. If we impose only $U_2\subset P_2$ then we
could allow ubereges of the type $\{v,h\}$ with $h\notin U_1$, thus
allowing metabolites to affect reactions which are not in the model.

\begin{example}
Given a set of three nodes $V=\{v_1,v_2,v_3\}$ we want to explore depth-$2$
ubergraphs. First $P_1=\mathcal{P}\left(V\right)\setminus \{ \emptyset \}$
thus 
\[P_1 = \{ \{v_1\},\{v_2\},\{v_3\},
\{v_1,v_2\},\{v_1,v_3\},\{v_2,v_3\}, \{v_1,v_2,v_3\} \}.\]
In other words $P_1$ contains all loops around nodes $v_i$,
regular edges and the only possible (strict) hyperedge $\{v_1,v_2,v_3\}$.
$U_1$ is any subset of $P_1$. The cardinality of $P_1$ is
$2^{3}-1=7$, since it is the power set of $V$ excluding the empty set.\\
$P_2= \mathcal{P}\left(V\cup P_1\right)\setminus \{ \emptyset \}$
contains all ubereges 
$\{v_{k_1},\ldots,v_{k_m},h_{i_1,},\ldots,h_{i_p}\}$, with $0\leq m\leq 3$,
$0\leq p\leq 3$, $m+p\geq 1$, $v_{k_l}\in V$ and $h_{i_l}\in P_1$.
$P_2$, being a power set excluding the empty set, has cardinality $2^{3+2^3-1}-1=2^{10}-1 = 1023$. Notice that there are objects not easily interpreted
in $P_2$, for instance the uberedge $\{ \{v_1\}\}$, that is the uberedge
which loops over the loop over $v_1$.\\
Finally, the possible ubergraphs on three nodes are given by
a triplet $V,U_1,U_2$. The possible choices of $U_1$ are among all
subsets of $P_1$ thus we do have $2^{7}=128$. The possible choices
for $U_2$, not taking into account $U_1$, are given by $2^{|P_2|}=2^{1023}$
(where we used $|\cdot|$ to indicate the cardinality of a set). 
However, as explained above, the admissible $U_2$ may be chosen only as subsets of  $\mathcal{P}\left(V\cup U_1\right)\setminus \{ \emptyset \}$
thus the possible choices are $2^{2^{3+|U_1|}-1}$. 
\end{example}

As shown above, the set of ubergraphs can be extremely complex
and of high cardinality. Moreover, not all uberedges can be easily interpreted.
Thus on one side we would like to restrict the definition to allows only
objects with modeling meaning, but on the other side a further generalization is
necessary to allow directed hper and uberedges, weights and signs.
This is achieved by next definition.
\begin{definition}
A depth-$k$ metabolic graph is an ubergraph with (nontrivial) directed weighted
hyperedges and (nontrivial) directed signed uberedges of depth greater than  or equal to $2$.
More precisely it is a $k+3$-tuple $G = (V,{\cal H},\Psi_{\cal H},U_2,\ldots,U_k,\Psi_{U})$
such that the following holds:
$V$ is a finite set; ${\cal H}$ is the set of directed hyperedges
$(X,Y)$, $X,Y\subset V$, $X,Y\not= \emptyset$;
$\Psi_{\cal H}=\{\Psi^{out}_h,\Psi^{in}_h\}_{h\in H}$,
$\Psi^{out}_h:X\to \R_+$, $\Psi^{in}_h:Y\to \R_+$,  is the set of weights;
for $j=2,\ldots,k$, $U_j={(X,Y)}$ with $X,Y \subset\mathcal{P}\left(\bigcup_{i=0}^{j-1}U_i\right)\setminus \{ \emptyset \}$, $X,Y\not= \emptyset$;
$\Psi_{U}=\{\Psi^j\}_{2\leq j\leq k}$ with $\Psi^j:U_j\to \{-1,+1\}$.
\end{definition}


\begin{example}
A depth-$2$ metabolic graph includes directed weighted hyperedges
and signed uberedges of the type $(X,Y)$ with $X,Y$ containing nodes
and hyperedges. Such a type of uberedge may represent the
enhancer or inhibitor action of a set of metabolites and reactions
(those in $X$) over another set of metabolites and reactions 
(those in $Y$).\\
A depth-$3$ metabolic graph may also include the action of
a depth-$2$ uberedge over another set of depth-$2$ ubderedges.
This type of edges may represent the effect of a part of a metabolic
network over another one by molecular mechanisms which are not known.
\end{example}

\clearpage 

\bibliography{TB_references_Archive}  
\end{document}

%% file: intro.tex
\begin{tikzpicture}[every node/.style={scale=1},->,shorten >=1pt,auto,node distance=1cm,
        thick,main node/.style={circle,fill=blue!10,draw,minimum size=0.5cm,inner sep=0pt]}]
\node[main node] at (-1,2.5) (1) {1};
\node[main node] at (-1.2,1.5) (2) {2};
\node[main node] at (-1,0.5) (3) {3};

\node[main node] at (0.9,2.2) (4) {4};
\node[main node] at (0.9,0.8) (5) {5};

\node[draw=none] at (-0.05,1.8) {\Large $h$};


	\path[-]
    (1) edge[bend right, ultra thick] node[xshift=-0.3cm,yshift=0.2cm] {$\alpha_1$} (0.15,1.5) ;
    \path[-]
    (2) edge[ultra thick] node[xshift=-0.3cm,yshift = -0.1cm] {$\alpha_2$} (0.15,1.5) ;
    \path[-]
    (3) edge[bend left, ultra thick] node[below,xshift = 0cm,yshift=-0.2cm] {$\alpha_3$} (0.15,1.5) ;
    \path[-]
    (0.1,1.5) edge[->,bend right, ultra thick] node[above,xshift = -0.1cm,yshift = 0cm] {$\alpha_4$} (.75,1.95);
    \path[-]
    (0.1,1.5) edge[->,bend left, ultra thick] node [below,xshift = -0.1cm,yshift = 0cm] {$\alpha_5$} (.75,1.05);

\end{tikzpicture} \hspace{2cm}  
\begin{tikzpicture}[every node/.style={scale=1},->,shorten >=1pt,auto,node distance=1cm,
        thick,main node/.style={circle,fill=blue!10,draw,minimum size=0.5cm,inner sep=0pt]}]
\node[main node] at (-1.2,1.5) (1) {1};

\node[main node] at (1,1.5) (2) {2};

\node[draw=none] at (-0.5,1.7) {\Large $e$};

\node[main node, rectangle, rounded corners=.2cm] at (0, 0.5)  (enhancer) { \;Enhancer\; };
\node[main node, rectangle, rounded corners=.2cm] at (0, 2.5)  (inhibitor) { \;Inhibitor\; };

    \path[-]
    (1) edge[->,ultra thick] node[xshift=-0.3cm,yshift = -0.1cm] {} (2);

    \draw[-o] (enhancer) -- (0,1.4);    
	\draw[-|] (inhibitor) -- (0,1.6); 
	 \node[draw=none] at (0.3,1.9) {$\left( - \right)$};
	 \node[draw=none] at (0.3,1.0) {$\left( + \right)$};
	 \node[draw=none] at (-0.2,2) {$u_1$};
	 \node[draw=none] at (-0.2,1.0) {$u_2$};
\end{tikzpicture}

%% file: EI.tex
To model intakes, excretions, general biochemical reactions, inhibitors and enhancers acting on a metabolic network, we will introduce 
a new mathematical object called a \textit{metabolic graph}.
Metabolice graphs are a subclass of ubergraph, a generalization of hypergraphs introduced in \cite{joslyn2017ubergraphs} (here we use a different definition of ubergraph stated in the Appendix).

Directed graph are commonly used to model metabolic networks, with nodes representing metabolites and edges representing the biochemical reactions. A hypergraph is a more general structure where an edge can connect more than two nodes. For the following definitions, let $V=\{v_1,\ldots,v_n\}$ be a set of nodes and let $\mathcal{P}(V)$ be the power set of $V$.

\begin{definition}
A \textit{hyperedge}  $h$ is a set of nodes connected to each other, i.e. 
$h\in\mathcal{P}(V)\setminus \{\emptyset\}$. Note that the set $h = \{v_i\}$ indicates a loop edge connecting $v_i$ to itself.
\end{definition}
Because chemical reactions of a metabolic network indicate a direction of flow, we require hypergraphs to include direction.
\begin{definition}
A \textit{directed hyperedge} is an ordered pair of two subsets of nodes, i.e. $h=(X,Y)$ with $X\in\mathcal{P}(V)$ or $X = \{v_0\}$, $Y\in\mathcal{P}(V \cup \{v_{n+1} \})$, where $v_0$ $(v_{n+1})$ is a virtual node called the source (sink). Elements of $X$ $(Y)$ are called initial nodes (terminal nodes) for the hyperedge $h$. The set of directed hyperedges is denoted ${\cal H}$. 
\end{definition}
When using a hyperedge to model reacting metabolites forming a product, we encode the stoichiometry relationship on the a hyperedge via edge weights. It is convenient to have notation for the cardinality of the sets of initial and terminal nodes for a hyperedge.
\begin{definition}
Given a directed hyperedge $h = (X, Y)$, with $X,Y\in\mathcal{P}(V)$, 
the \emph{indegree} of $h$ is defined as
\begin{equation}
d_{\text{in}}(h) =  |X|,
\end{equation}
and the \emph{outdegree} of $h$  is defined as
\begin{equation}
d_{\text{out}}(h) =  |Y|
\end{equation}
where $|\cdot|$ indicates the cardinality of a set.
\end{definition}
\begin{definition}
A weighted directed hyperedge is a couple ${\cal H}\ni h=(X,Y)$ with \
$X\in\mathcal{P}(V)$ or $X = \{v_0\}$, $Y\in\mathcal{P}(V\cup \{v_{n+1}\})$, and corresponding weights $\Psi_h: h \mapsto (\Psi_h^{\text{out}},\Psi_h^{\text{in}})$ where
$\Psi_h^{\text{out}}:X\mapsto \mathbb{R}_+$ and $\Psi_h^{\text{in}}:Y\mapsto \mathbb{R}_+$. 
\end{definition}
We are now ready to describe a mathematical representation
of inhibitors and enhancers. More precisely, we want to consider metabolites influencing a given reaction. 
This can be captured by new type of generalized edges, called an \textit{uberedge} linking nodes to directed hyper edges.  
\begin{definition}
An e/i-uberedge is a couple $u=(v,h)$ with $v\in V, h\in {\cal H}$. We denote the set of 
e/i-uberedges by ${\cal U}$.
\end{definition}
We are now ready to provide the definition of metabolic graph.
For a complete description of a metabolic networks, uberedges are endowed with a sign to indicate their action as enhancer $(+)$ or inhibitor $(-)$. This subset of uberedges from a node to a hyperedge are depth 2 uberedges according
to the general definition of ubergraph (see Appendix).
\begin{definition} A metabolic graph is a weighted directed hypergraph endowed with signed depth-2 uberedges connecting nodes to hyperedges. More precisely, a metabolic graph is an ordered quintuplet 
$G=(V,{\cal H},{\cal U},\Psi_{\cal H},\Psi_{\cal U})$ where $V=\{v_1,\ldots,v_n\}$ 
is the set of nodes, ${\cal H}$ is the set of directed hyperedges, $\Psi_{\cal H}=\{\Psi_h: h\in {\cal H}\}$ is the set of functions assigning weights to hyperedges, 
${\cal U}$ is the set of e/i-uberedges and $\Psi_{\cal U}: {\cal U}\mapsto \{+,-\}$.
\end{definition} \label{def:metabolic graph}
\begin{definition}
Given a metabolic graph $G$,
a \emph{path} is a sequence of distinct nodes $v_{i_1}\cdots v_{i_k}$, with $(v_{i_j}\in X,v_{i_{j+1}}\in Y)$ and $(X,Y)\in {\cal H}$ for $j=1,\ldots,k-1$.
A graph is \emph{strongly connected} if there exists a path between every pair of nodes. A \emph{strongly connected component} of a directed graph is a maximal strongly connected subgraph.\\
A \emph{terminal component} of a metabolic graph $G$ is a strongly connected component corresponding to a subset of nodes $V'\subset V$, 
such that for  $v'\in V'$ and $v\in V\setminus V'$ there exists no hyperedge $h=(X,Y)$ with $v'\in X$ and $v\in Y$.
\end{definition}
\begin{definition}
Given a metabolic graph $G$,
\emph{Intakes} (\emph{Excretions}) are hyperedges $(X,Y)$ such that $X = \{v_0\}$ ($v_{n+1} \in Y$).  \emph{Intake nodes} are nodes $v\in V$ such that there exits a hyperedge $(X,Y)$ with $X = \{v_0\}$ and $v\in Y$. \emph{Excretion nodes} are nodes $v\in V$ such that there exits a hyperedge $(X,Y)$ with $v\in X$ and $v_{n+1} \in Y$. We indicate by $I$ the set of intake nodes and by $J$ the set of excretion nodes.
\end{definition}

\subsection{Central Carbon Metabolism}

Let us go back to the central carbon metabolism network of MTB represented in Figure \ref{fig:central_carbon}.
This network can be described by a metabolic graph $G=(V,{\cal H},{\cal U},\Psi_{\cal H},\Psi_{\cal U})$
as follows. The set $V$ of nodes is given by:\\
$V= \{ v_1,\ldots, v_{11} \}=
\{\mbox{glucose}, \mbox{PEP}, \mbox{pyruvate}, \mbox{acetyl CoA}, \mbox{isocitrate}, \mbox{oxaloacetate},\\ \mbox{glyoxylate}, \mbox{malate},
\alpha-\mbox{ketoglutarate}, \mbox{succinate}, \mbox{Clofazimine}\} $.\\ The set of hyperedges ${\cal H}$
contains regular edges and three directed hyperedges:\\
$h_1=(\{\mbox{acetyl CoA},\mbox{oxaloacetate}\},\{\mbox{isocitrate}\})$,\\$h_2=(\{\mbox{isocitrate}\},\{\mbox{acetyl CoA},\mbox{oxaloacetate}\})$,\\
and $h_3=(\{\mbox{acetyl CoA},\mbox{glyoxylate}\},\{\mbox{malate}\})$.\\
The set of uberedges ${\cal U}$ has only one uberedge:\\ 
$u_1=(\mbox{Clofazimine},(\{ \mbox{malate}\},\{ \mbox{oxaloacetate}\}))$.\\ 
For simplicity the functions $\Psi_{\cal H},\Psi_{\cal U}$ are not listed,
but can be deduced from the KEGG database \cite{kanehisa2000kegg}.\\
This graph is more descriptive of important functions of drug action on the metabolic network, and it contains three hyperedges and a single uberedge. Each of the hyperedges combines three metabolites; $h_1$,$h_3$ have two reactants and one product whereas $h_2$ has two reactants and one product. 
The uberedge shown from Clofazimine to the malate-oxaloacetate reaction indicates the well known TB drug ``Clofazimine'' acts as an inhibitor on a reaction which oxidizes malate.

\section{Metabolic dynamics with inhibitors and enhancers}\label{sec:EI}
In the treatment of TB, drugs will act as enhancers and inhibitors to various edges in the network, and so will appear as uberedges. In this section we define how enhancers and inhibitors affect the dynamics as well as equilibria conditions. 
In order to include enhancers and inhibitors (which serve as the initial nodes for uberedges) we introduce a new assumption:\\

\noindent
{\bf (H2)} We assume that for every node $v$ and hyperedge $h=(X,Y)$, with $v\in X$,
the following holds. Let $U_h$ be the set of nodes $w$ such that there exists e/i-uberedge $(w, h)\in {\cal U}$, then we have:
\begin{equation} \label{eq:S_construction}
S_{v h}(x) = \begin{cases}
 -\alpha_v\cdot {\bf F}_h(x)\cdot {\bf K}_h(x) & v\in X\\
 \alpha_v\cdot {\bf F}_h(x)\cdot {\bf K}_h(x) & v\in Y\\
 1 & X=\{v_0\}, v\in Y,\\
0 & \mbox{otherwise},
\end{cases}
\end{equation}
where $\alpha_w=\Psi_h^{\text{in}}(w)$ if $w\in X$ and
$\alpha_w=\Psi_h^{\text{out}}(w)$ if $w\in Y$ are the stoichiometric coefficients,
${\bf F}_h:\mathbb{R}^{d_{in}(h)}\to \mathbb{R}_+$ is given by
\begin{equation}
{\bf F}_h(x) = \min_{w\in X} \left\{F_{w,h} (x_w) \cdot \frac{1}{\alpha_w}\right\},
\end{equation} 
$F_{w,h}:\R_+\to \R_+$ quantifies the potential flow of metabolite $x_w$
due to reaction $h$, and
\begin{equation}
{\bf K}_h=\prod_{w\in U_h} K_{(w,h)}(x_w),
\end{equation}
where $K_{(w,h)}:\mathbb{R}_+ \to \mathbb{R}_+$ quantifies the action of metabolite $x_w$ on $h$, 
with the convention that $K_h=1$ if $U_h=\emptyset$.\\
The stoichiometric coefficients $\alpha_w\in\mathbb{R}$ for the reaction corresponding to hyperedge $h$are normalized such that $\sum_{w \in X} \alpha_w=1$ and $\sum_{w \in Y} \alpha_w=1$.
The functions $F_{w,h}$, $w\in Y$ are continuously differentiable.
The functions $K_{(w,h)}$, $w\in U_h$, are continuously differentiable, monotonic with $K_{(w,h)}(0)=1$.
More precisely if $\Psi_U((w,h))=+$ then
$K_{(w,h)}$ is increasing (enhancer case), othwerwise 
$K_{(w,h)}$ is decreasing (inhibitor case).\\

\noindent
Similarly to (H1), under assumption (H2) each function $F_{w,h}$ depends only  on the metabolite $x_w$, but there is the additional factor ${\bf K}$ which corresponds to the action of one or more e/i-uberedges. This gives a nonlocal dependence, with respect to network topology, because the node(s) corresponding to an enhancer(s) or inhibitor(s) 
may be anywhere in the network not necessarily close to the edge it is affecting.\\
We are ready to state our first result:
\begin{proposition}\label{prop:A_neccessary}
Consider a system \eqref{eq:first-dyn-LIFE} satisfying (H2).
Assume there exists an equilibrium $\bar{x}\in (\mathbb{R}_+)^n$ 
for a flux vector $f$ such that $f_h>0$ for every $h\in {\cal H}$.
Then for every node $v \in V$ for which there exists a path from the intake nodes to $v$,  there exists a path from $v$ to the excretion nodes.
\end{proposition}
\begin{proof}
Assume there exists an equilibrium $\bar{x}\in \mathbb{R_+}^n$ and, by contradiction, a node $v$ for which there exists a path from $w$, an intake node, to $v$, but there exists no path from $v$ to some excretion node. Since there is no path from $v$ to excretion nodes, either $v$ belongs to a terminal component, or there is a path from $v$ to a terminal component with no excretion. Denote by $V_T\subset V$ the set of nodes of such a terminal component. Since there is a path from $w$, an intake node, to $v$ and a (possibly trivial) path from $v$ to $V_T$, then there is also a path from intake nodes to $V_T$. Denote by $v_0, v_1 = w, \dots, v_{\ell-1},  v_{\ell}$  one such a path, such that $v_{\ell-1} \notin V_T$ and $v_{\ell}\in V_T$ (possibly the path is a single hyperedge, in the case with $w \in V_T$).\\
It is easy to show that $x_{v_i} >0$ for all $i=1, \dots, \ell$, as follows.
If $h=(X,Y)$ with $v_0\in X, v_1\in Y$, then by (H2) we have $S_{v_0, h}(\bar{x})=1$.
On the other side, for every $h'=(X,Y)$ with $v_1\in X, w'\in Y$, $x_{v_1}=0$ implies $S_{v_1,h'}(x)=0$. 
Consequentially $x_{v_1}=0$ implies $\dot{x}_{v_1} \ge \alpha_{v_1}f_h {\bf F}_h(\bar{x})>0$ (where $h=(X,Y)$ with $v_0\in X, v_1\in Y$), contradicting $\bar{x}$ being an equilibrium. 
Having proved that $\bar x_{v_1}>0$, we can proceed by induction: for $i=1, \dots, \ell-1$,  $\bar x_{v_i}>0$ implies $\bar x_{v_{i+1}}>0$. The argument is the same as above, with a slight modification: looking at $h=(X,Y)$ with $v_{i}\in X, v_{i+1}\in Y$, $S_{v_{i+1}, h}(\bar x) = \alpha_v\cdot {\bf F}_h(\bar{x})\cdot {\bf K}_h(\bar{x})>0$
thanks to (H2) together with $\bar x_i>0$, while above we were in the case of an intake.\\
Finally we have a terminal component with no excretion, and a hyperedge $h=(X,Y)$ with $v_{\ell-1}\in X, v_{\ell} \in Y$ with $v_\ell \in V_T$ and $v_{\ell-1}\notin V_T$, such that either $v_{\ell-1} = v_0$ or $\bar x_{v_{\ell-1}}>0$. In either case, considering ${\tilde{h}}=(X,Y)$ with $v_{\ell-1}\in X, v_\ell\in Y$, by (H2) we have $S_{v_{\ell},{\tilde{h}}}(\bar x) = \alpha_{v_{\ell}}\cdot {\bf F}_{\tilde{h}}(\bar{x})\cdot {\bf K}_{\tilde{h}}(\bar{x})>0$.
Now consider the variation of mass in the nodes of the component $V_T$:
since there are no hyperedges leaving $V_T$, and there is at least the incoming hyperedge ${\tilde{h}}$, we have
$ \frac{\mathrm d}{\mathrm d t} \sum_{v\in V_T} x_v
=  \sum_{v\in V_T} \dot x_v 
\ge  \alpha_{v_{\ell}}\cdot {\bf F}_{\tilde{h}}(\bar{x})\cdot {\bf K}_{\tilde{h}}(\bar{x}) f_{\tilde{h}} >0$, contradicting the fact that $\bar{x}$
is an equilibrium. 
\qed \end{proof}

\noindent
For a system with fixed metabolites, the existence of feasible flows will depend on network structure.
The Max-flow-min-cut Theorem \cite{ford1956maximal} implies that a feasible flow exists if there is a path from intakes to excretions, and indeed this Theorem holds true also for metabolic graphs providing suitable definitions.

\begin{definition}\label{def:flow-MG}
A flow on a metabolic graph $G=(V,{\cal H},U,\Psi_{\cal H}, \Psi_U)$ is a function $g:{\cal H}\mapsto R_+$ such that $g$ satisfies Kirchhoff's law for metabolic graphs, i.e. for every node $v$:
\begin{equation}\label{eq:Kirchhoff}
\sum_{h\in \Gamma^{out}(v)} \Psi_h^{out}(v)\cdot g(h)=\sum_{k\in \Gamma^{in}(v)} \Psi_k^{in}(v)\cdot g(k)
\end{equation}
where $\Gamma^{out}(v)= \{(X,Y)\in {\cal H}:v\in X\}$, $\Gamma^{in}(v)= \{(W,Z)\in {\cal H}:v\in Z\}$.
\end{definition}

\begin{definition}
For a given flow $g$ on a metabolic graph $G$, the \emph{amount of flow} from $v_0$ to $v_{n+1}$ is 
$v(g)= \sum_{h\in {\cal H},v_0\in X} g(h) = \sum_{h\in {\cal H},v_{n+1}\in Y} \Psi_h^{out}(v_{n+1})\cdot g(h)$. 
\end{definition}

\noindent
\textbf{Maximal Flow Problem.} Consider a metabolic graph with a function $c:{\cal H}\mapsto R_+$ assigning to each edge a maximal capacity. The maximal flow problem is defined
\begin{equation}
\max(v(g)) \text{ such that } g(h) \leq c(h) \text{ for every } h \in {\cal H}.
\end{equation}

\noindent
We are now ready to define a cut.
\begin{definition}
Consider a metabolic graph $G=(V,{\cal{H}},{\cal{U}},\Psi_{\cal{H}}, \Psi_{\cal{U}})$.
Given $S,T \subset V$, we define ${\cal H}(S,T) = \{(X,Y)\in {\cal H} :X \cap S \neq \emptyset$ 
and $Y \cap T \neq \emptyset\}$, as the set of edges connecting nodes of $S$ to nodes of $T$.\\ 
Given a flow $g$ on $G$, the total flow from nodes in $S$ to nodes in $T$ is defined by:
\[
g(S,T)= \sum_{h=(X,Y) \in {\cal H}(S,T), i\in Y \cap T} \alpha_i h.
\]
\end{definition}
\begin{definition}
Consider a metabolic graph $G=(V,{\cal{H}},{\cal{U}},\Psi_{\cal{H}}, \Psi_{\cal{U}})$ with source $v_0$, and sink $v_{n+1}$. 
Let $S \subset V \cup v_0$ be a set such that $v_0 \in S$ and $v_{n+1} \notin S$ and 
define $T=(V \cup \{v_{n+1}\})   \setminus S$. 
Then the set of hyperedges $C_S={\cal H}(S,T)$ is called a \emph{cut separating $v_0$ from $v_{n+1}$}. 
The capacity of the cut is defined by:
\begin{equation}
c(C_S) = \sum_{h \in C_S}  c(h).
\end{equation}
Notice that the capacity of the branch of hyperedge $h$ adjacent to a node $v$ can be written as $c(h) \cdot \alpha_v$.
\end{definition}

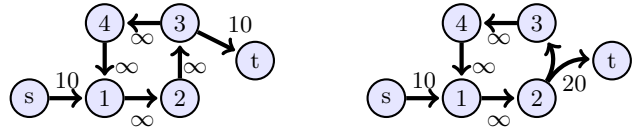
\begin{figure}[h!]
\hspace{2.3cm}\input{Maxflow_example_graph.tex}
\caption{\footnotesize A directed graph (left) and a metabolic graph (right) with similar structure. Capacities are listed on edges and the hyperedge has a 1:1 ratio of flow for the two branches, i.e., 10 units flow across each branch. The maximum flow for each graph is 10, however the metabolic graph has a unique solution while the directed graph does not. }\label{fig:maxflowgraphs} 
\end{figure}

The max flow of a directed graph is not necessarily unique. For a graph with a directed cycle, the flow on the cycle can be increased (capacity permitting) without violating Kirchoff's law, or changing the total flow through the graph. For example, the directed graph in Figure \ref{fig:maxflowgraphs} (left) admits the maximal flow defined by setting the flow on edges (s,1),(1,2),(2,3) and (3,t) equal to 10 while the flow on (4,1),(3,4) equal to 0. However, the maximal flow can also be achieved setting (s,1),(3,4),(4,1) and (3,t) to 10 while the flow on (1,2),(2,3) to 20. Often when computing maximum flow on directed graph the flow through cycles is eliminated. In a metabolic graph with hyperedges, flow through cycles may be unavoidable. The metabolic graph in \ref{fig:maxflowgraphs} (right) contains one hyperedge, with branches that are equally weighted, one of them leading to a sink, while the other leads to a cycle. This network admits a unique maximum flow. This is produced by setting (s,1),(3,4),(4,1) to 10 while the flow on (1,2),(2,(3,t)) to 20. Note that setting the flow (2,(3,t)) to 20 will cause 10 to flow to t and 10 to flow to 3 due to this hyperedge having equal weighting to its branches. 

\noindent
We are now ready to state the following:
\begin{lemma}\label{lemma:flowcut}
Given a flow $g$ from $v_0$ to $v_{n+1}$, and a cut $C_{S}$ separating $v_0$ from $v_{n+1}$, it holds:  $v(g) \leq c(C_S)$
\end{lemma}
\begin{proof}
Fix a flow $g$ and a cut $C_{S}$ separating $v_0$ from $v_{n+1}$. The amount of flow $v(g)$ is equal to the flow entering the graph, equivalently the amount of flow entering $S$. We also have $v(g)$ will equal the transfer of flow from $S$ to $T$,
\begin{equation*}
v(g) = g(S,T) - g(T,S).
\end{equation*}
Since flows are positive we have
\begin{equation*}
g(S,T) - g(T,S) \leq g(S,T)=\sum_{h=(X,Y) \in C_{S}, i\in Y \cap T} \alpha_i h \leq \sum_{h\in C_{S}} h = c(C_{S}).
\end{equation*} 
 \qed \end{proof}
Since Lemma \ref{lemma:flowcut} applies to every flow $g$ and cut $C_{S}$, we can take the maximum over all flows and the minimum over all cuts to obtain
\begin{equation}\label{eq:flowlesscut}
\max(v(g)) \leq min(c(S)).
\end{equation}
The following Proposition shows that, as for simple graphs, an equality holds in \eqref{eq:flowlesscut}.
\begin{proposition}
(Max-flow-min-cut for Metabolic graphs)
Formula \eqref{eq:flowlesscut} holds with equality sign, i.e.:
$$  \max(v(g)) = \min(C_{S}) $$
\end{proposition}

\begin{proof}
Take a flow $g$ such that $v(g)$ is maximal. We construct a cut starting from the intake nodes by recursion. Define $S_0 =I$ to be all intake nodes of the metabolic graph. At each step $\nu$, we define $S_{\nu+1}$ from $S_{\nu}$ as follows: 
\begin{enumerate}
\item If $v \in S_{\nu}$ then $v \in S_{\nu+1}$,
\item if $v_1,v_2\in S_{\nu},w \in V$ and there exists a hyperedge $h=(X,Y)$ such that $v \in X,w \in Y$ and $g(h)<c(h)$ or $w\in X,v_2\in Y$ and $g(h)>0$,then $w\in S_{\nu+1}$,
\item if $v_1,v_2\in S_{\nu}$ and there exists a hyperedge $h=(X,Y)$ such that $v_1 \in X, v_2 \in Y$, then for all $w\in Y$, $w\in S_{\nu+1}$
\end{enumerate}
Clearly there will only be a finite number of steps until $S_\nu$ is stable, i.e. no more edges will be added after the $\bar{\nu}-th$ step. We define $S= S_{\bar{\nu}}$ and $T={V} \setminus S$.\\
We claim that $v_{n+1} \notin S$. Assume, by contradiction that there exists a path 
$v_{i_1},...,v_{i_l}$ such that $v_{i_1}=v_0$, $v_{i_l} = v_{n+1}$. 
Then define for $k = 1,...,l-1$,
\begin{equation*}
\epsilon_k = \max\left\{\sum_{h =(X,Y) : v_{i_k} \in X,v_{i_{k+1}} \in Y } (c(h) - g(h)),\sum_{h =(X,Y) : v_{i_{k+1}} \in X,v_{i_k} \in Y } g(h)\right\}>0
\end{equation*}
and $\epsilon= \min_k \epsilon_k$. We now define a new flow $\tilde{g}$ to reach a contradiction as follows. 
If 
$\epsilon_k = \sum_{h =(X,Y): v_{i_k} \in X,v_{i_{k+1}} \in Y} (c(h) - g(h))$, then for each $h \in K_1$ we set 
\begin{equation*}
\tilde{g}(h) = g(h)+\epsilon. 
\end{equation*}
Otherwise 
we set
\begin{equation*}
\tilde{g}(h) = g(h)-\epsilon. 
\end{equation*}
By construction $\tilde{g}$ is a flow, moreover $v(\tilde{g}) = v(g) + \epsilon $. Then we reach a contradiction 
for the maximality of $v(g)$.\\
Define the cut $C_S = {\cal H}(S,T)$. Then the flow from $v_0$ to $v_{n+1}$ satisfies $v(g)= g(S,T)-g(T,S).$ 
Since $g$ is maximal, we have $g(T,S)=0$ and $g(S,T)=\sum_{h\in C_S} (g(h))=\sum_{h\in C_S} c(h) = c(C_S)$. Therefore we have $v(g)=c(S)$.   
\qed \end{proof}

\begin{definition}
Given a metabolic graph with intakes, the vector $\bar{\phi}$ represents the intake flows to each node, i.e. $\bar{\phi}_i=f_{h(v_0,v_i)}$ if $v_i$ is an intake node and $\bar{\phi}_i=0$ otherwise. 
\end{definition}

\begin{proposition}\label{prop:feasibleflows}
Given a system satisfying (H2), fix an $x \in (\mathbb{R}_+)^n$ and intake flow vector $\bar{\phi}$ with strictly positive entries.
There exists $f \in (\mathbb{R}_+)^{m}$ in the null space of $S(x)$ if and only if for each intake there exists a path to an excretion.
\end{proposition}

\begin{proof}
Consider the maximum flow problem on the metabolic graph $G$, where intake edges $h_I$ have capacity $\bar{\phi}$, and all other edges have infinite capacity.
The feasible flows $\varphi$ for this network are in one-to-one correspondence with the equilibrium fluxes $f \in \mathcal{N}(S(x)) \cap (\mathbb{R}_+)^{m}$ (where $\mathcal{N}(S(x))$ indicates the kernel of the matrix $S(x)$)
such that $f_h \le \bar{\phi}$ for all $h \in h_I$. The correspondence is simply given by $f_h  = \varphi_h$ for all $h\in h_I$ and $f_h  = \varphi_h / F_h(X) $ for all $h=(X,Y) \in {\cal H} \setminus h_I$.\\
If for all $h\in h_I$ there is a path to an excretion, then the minimum cut is the collection of all edges $h\in h_I$. The maximum flow $\varphi^*$ then satisfies $\varphi^*(h_I) = \bar{\phi}$, thus also ensuring the existence of an equilibrium flux $f^*$ satisfying the same property.\\
If for some $h\in h_I$ there is no path to an excretion, then all feasible flows $\varphi$ satisfy $\varphi_h  = 0$, and hence all equilibrium fluxes $f^*$ satisfy $f^*_h = 0$ which contradicts the assumption.
\qed \end{proof}


\begin{figure}[h!]
\begin{center}
\hspace{2.3cm}\input{Counter_example_graph.tex}
\caption{\footnotesize In this metabolic graph node 2 acts as an inhibitor to an edge leaving itself.}\label{fig:counterexample} 
\end{center}
\end{figure}
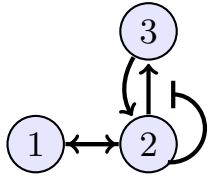
\vskip -1cm
\noindent
In \cite{maeda1978asymptotic} several Propositions guarantee boundedness of solutions and no periodic oscillation.  Specifically, the main results are: (1) Trajectories are bounded if and only if there exists an equilibrium with positive entries; (2) If trajectories are bounded, then they approach an equilibrium set for $t \to \infty$. 
These conclusions are also true under assumption (H1) for simple graphs, but not under assumption (H2) for metabolic graphs.\\
We start providing a counterexample via a small system of three nodes and one inhibitor, see Figure\ref{fig:counterexample}. 
We assume (H2) to hold with $F(x_v) =x_v$ for all nodes $v$, 
\begin{equation*}\label{eq:K2}
K_2(x_2) = \begin{cases}
 \frac{1}{3}((x-6)^2+5) & x\leq 6 \\
      \frac{5}{3} & 6< x, \\
\end{cases}
\end{equation*}
and stoichiometric matrix given by
\begin{equation*}\label{eq:S_example}
S= \left(\begin{matrix}-x_1&0&x_2&0\\x_1&-x_2 \cdot K_2(x_2)&-x_2&x_3\\0& x_2 \cdot K_2(x_2)&0&-x_3\end{matrix}\right), 
\end{equation*}
with flux vector $f = [1,3,3,2].$ The structure is visualized in Figure \ref{fig:counterexample}.
This example will admit an oscillatory solution because $x_2 \cdot K_2(x_2)$ is not monotone increasing and has a negative slope for $\frac{1}{3} (12 - \sqrt{21})<x_2<\frac{1}{3} (12 + \sqrt{21})$. The example reduces to the example given in Remark 2 of \cite{maeda1978asymptotic}, where further details can be found.

\noindent
Under assumption (H1) we have several additional conditions which relate to existence and uniqueness of equilibria. These results again depend on the structure of the network as well as an additional constraint on the functions $F_h$.
Adapted from \cite[Theorems~4 and 5]{maeda1978asymptotic} these are:
(1) There exists an equilibrium with positive entries for constant intakes if and only if for all $v \in V$ there is a path to $X$ such that all edges in the path have $\lim_{x_v\to \infty} H_e(x_v) = +\infty$;
(2) If there exists an equilibrium with positive entries, and all $v \in V$ connect to $J$, then the equilibrium is unique.
Even if we are able to relate the existence of equilibria to network structure via Proposition \ref{prop:A_neccessary} and \ref{prop:feasibleflows}), uniqueness of equilibria may fail without further assumption.
Moreover, computation of equilibria is highly nontrivial. Indeed, 
for a simple graph with a single enhancer or inhibitor, we can write the dynamics as follows
\begin{equation}\label{eq:dynamics_EI}
\dot{x}=J_1(f)\eta(x)+J_2(f) k(x) + \phi
\end{equation}
where $J_1$ is the Jacobian matrix without the edges affected by inhibitor/enhancer, $J_2$ is a matrix with the edges affected by inhibitor/enhancer, $\eta_i(x) = F_{v_i}(x_{v_i})$, $\phi_i = f_{(v_0,v_i)}$ if ($v_0,v_i) \in E$, and $\phi_i = 0$ otherwise, and $k_i(x) = K_{v_j}(x_{v_j}) \cdot F_{v_i}(x_{v_i})$ if $v_j$ acts on an edge starting from $v_i$.
If another node acts as an inhibitor, as it happens for the addition of drug to the system, the size of $J_1,J_2$ would be adjusted accordingly. 
Clearly (\ref{eq:dynamics_EI}) in general does not have a unique solution
and the computation of solutions
can not be done analytically.\\

\subsection{Exploring the space of equilibria with inhibitors and enhancers}\label{sec:EQ}
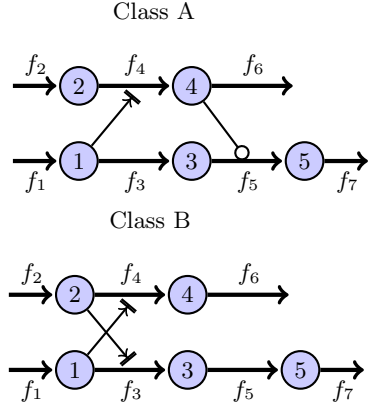
\begin{figure}[h]
\begin{center}
\input{double_example.tex}
\caption{\footnotesize Two networks with inhibitors and enhancers. Class A (left): inhibitor\slash enhancers behave sequentially: node $1$ affects downstream $f_4$ and so on. 
Class B (right), inhibitors\slash enhancers act in parallel.}\label{fig:double_example} 
\end{center}
\end{figure}

In this Section we further explore the problem of uniqueness and stability of equilibria for networks
with enhancers and inhibitors. First, we provide an example of non-uniqueness and
a condition to ensure uniqueness. Let us define two classes of systems:\\
{\bf Class A.} Enhancers and inhibitors act
in cascade, see Figure \ref{fig:double_example} top.\\
{\bf Class B.} Enhancers and inhibitors act in parallel, 
see Figure \ref{fig:double_example} bottom.

\noindent
To precisely define Class A and Class B networks let us first define an \emph{uberpath} and \emph{strictly upstream} on a metabolic graph $G=(V,{\cal H},{\cal U},\Psi_{\cal H}, \Psi_{\cal U})$.
\begin{definition} An \textit{uber-path} is a sequence of nodes $(v_{k_1}, \ldots, v_{k_m})$ such that for every $i=1,\ldots,m-1$ there is an hyperedge $(X,Y)\in {\cal H}$ for which one of the following holds:
1) $v_{k_i}\in X$, $v_{k_{i+1}}\in Y$; 2) $(v_{k_i},(X,Y))\in U$, $v_{k_{i+1}}\in X\cup Y$; 
3) $(v_{k_{i+1}},(X,Y))\in {\cal U}$, $v_{k_{i}}\in X\cup Y$.
\end{definition}
\begin{definition}
A node $v_0$ is \textit{upstream} of node $v_1$ if there exists a hyperpath from $v_0$ to $v_1$. 
A node $v_0$ is \textit{strictly upstream} of node $v_1$ if $v_0$ is upstream of $v_1$ and there is no uberpath from $v_1$ to $v_0$.  
\end{definition}
\begin{definition}
A network is called a \emph{Class A} if: 1) there exists a sequence of enhancers or inhibitors $(v_{k_1}, \ldots, v_{k_m})$ such that $v_{k_i}$ is strictly upstream of $v_{k_j}$ for all $i<j$; 2) all enhancers (and inhibitors) are strictly upstream of nodes connected to hyperedges that they enhance (inhibit). Precisely, part 2. means if there exists e/i-uberedge $u=(v_j,h^*)$ with $h^*=(X,Y)$ then $v_j$ must be strictly upstream of every node $v\in X\cup Y$.\\    
Networks that are not Class A are called \emph{Class B} networks.
\end{definition}
The conditions of Class A allow the equilibrium state of $v_1\in I$ to be determined based on the inflow and outflow for $v_1$.  The magnitude of the edge affected by $v_1$ at equilibrium state can then be calculated. This will allow $v_1$ to be determined at equilibrium.  Class A networks allow the equilibrium states of all enhancers or inhibitors to be calculated this way, which can determine all fluxes (even those affected by uberedges) in the system. Once all fluxes are determined, then the steady state of every node in the network can be calculated for this a unique steady state. This is shown in next:
\begin{proposition}
For a Class A network with a given flux, if there exists an equilibrium with positive entries, and all $v \in V$ connect to $J$, then the equilibrium is unique.
\end{proposition}
\begin{proof}
Let $G$ be a Class A metabolic graph.  Then there exists a sequence of $n$ enhancers or inhibitors $(v_{k_1}, \ldots, v_{k_m})\subset V$ with $v_{k_i}$ is strictly upstream of $v_{k_{i+1}}$, and therefore all nodes upstream of $v_{k_1}$ are unaffected by enhanced edges. It follows that $x_{k_1}$ has a unique steady state determined by fluxes corresponding to incoming and outgoing edges. According to $(H2)$: 
\begin{equation}
\dot{x}_{k_1}=\sum_{h\in \Gamma^{in}(v_{k_1})}\Psi_h^{\text{in}}(v_{k_1}){\bf F}_h(x)f_{h}-\sum_{h\in \Gamma^{out}(v_{k_1})}\Psi_h^{\text{out}}(v_{k_1}){\bf F}_h(x)f_{h}.
\label{eq:proposition}
\end{equation}
where we used the notation of Definition \ref{def:flow-MG}.
If for all $(X,Y) \in\Gamma^{in}$ we have $X\subset I$ then 
for all $v_i\in X$ we can write $\dot{x}_i =f_{j_0}-\sum_{h\in \Gamma^{out}(v_{k_1})}\Psi_h^{\text{out}}(v_{i}){\bf F}_h(x)f_{h}$. This determines the steady state for $x_i$, hence all steady state quantities in equation \eqref{eq:proposition} are known and the unique steady state of $x_{k_1}$ is determined. The same can
be obtained by recursion if the nodes $v_i$ are connected to $I$ by some
path with hyperedges not affected by enhancers or inhibitors,
which is always the case because $v_{k_1}$ is the first node acting
as enhancer or inhibitor. Finally the value of $x_{k_1}$ is determined.\\
Similarly, $v_{k_2}$ may be the terminal node, or initial node for some edges that are not affected by inhibitors or enhancers except possibly $v_{k_1}$, 
for which the value of $x_{k_1}$ has been determined. Denote the steady state value of $x_{k_1}$ as $\bar{x}_{k_1}$
According to $(H2)$,
either $\dot{x}_{k_2}=\sum_{h\in \Gamma^{in}(v_{k_2})}\Psi_h^{\text{in}}(v_{k_2}){\bf F}_h(x)f_{h}-\sum_{h\in \Gamma^{out}(v_{k_2})}\Psi_h^{\text{out}}(v_{k_2}){\bf F}_h(x)f_{h}$ or $\dot{x}_{k_2}=\sum_{h\in \Gamma^{in}(v_{k_2})}\Psi_h^{\text{in}}(v_{k_2}){\bf F}_h(x)K(\bar{x}_{k_1})f_{h}-\sum_{h\in \Gamma^{out}(v_{k_2})}\Psi_h^{\text{out}}(v_{k_2}){\bf F}_h(x)K(\bar{x}_{k_1})f_{h}$, with enhancer terms $K(\bar{x}_{k_1})$ inserted. Therefore the value of $x_{k_2}$ is also determined.\\
By recursion, we can determine the value of all enhancers and inhibitors nodes, and therefore, all enhanced or inhibited fluxes are determined.Finally we can determine uniquely the value of all metabolites and fluxes.
\qed \end{proof}

\begin{figure}[h]
\begin{center}
\includegraphics[scale=0.5, trim={5cm 9.85cm 5cm 10.285cm},clip]{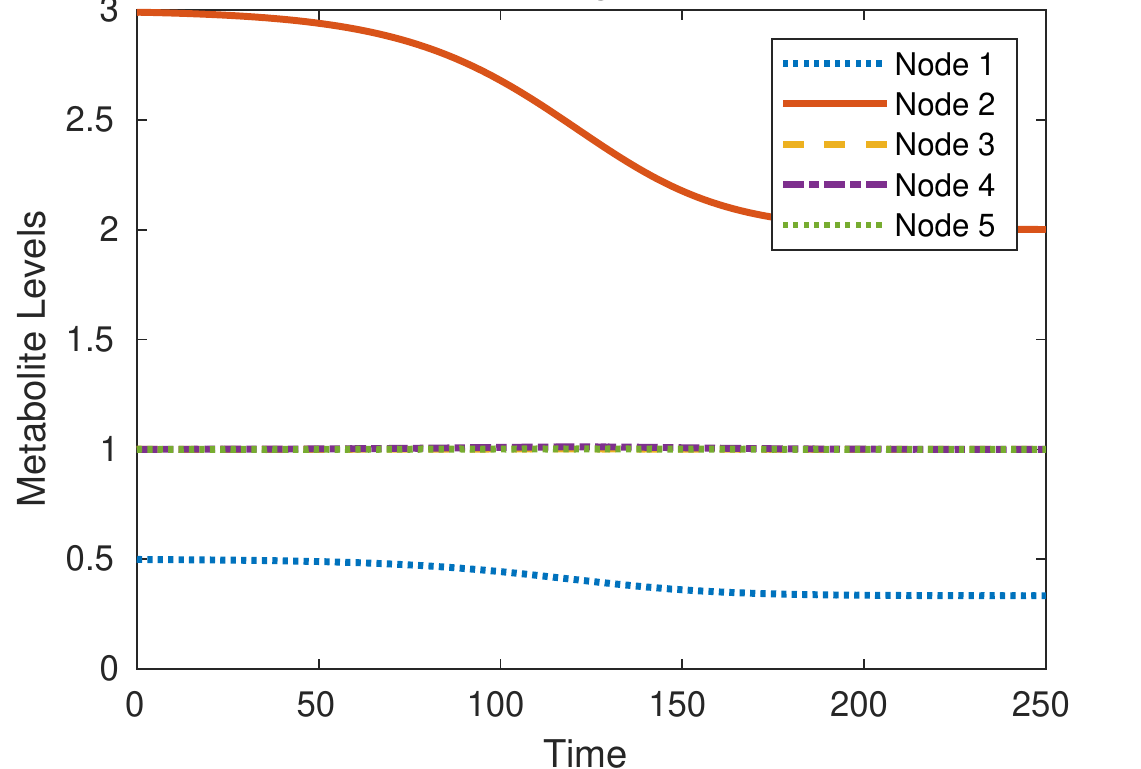}
\includegraphics[scale=0.51, trim={4.8cm 10cm 5.5cm 10cm},clip]{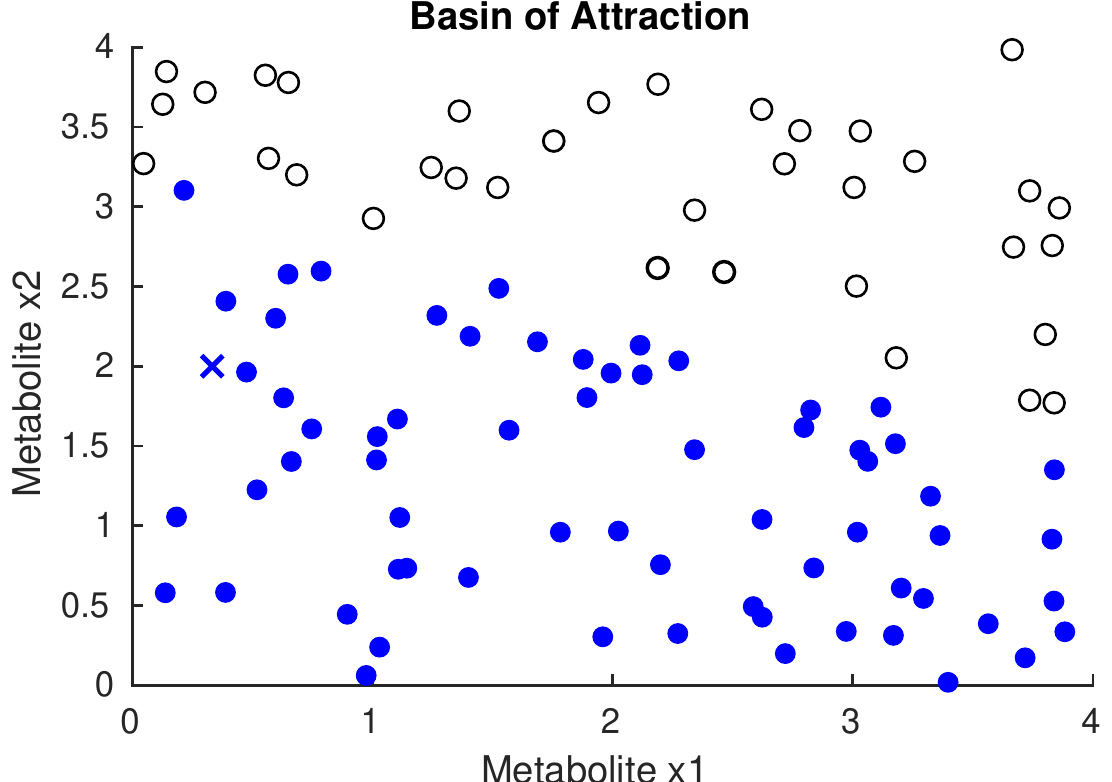}
\end{center} 
\caption{\footnotesize Simulations for the class B system in Fig. \ref{fig:double_example}. Top: Starting near the unstable equilibrium $\tilde{x}$, the system tends to the stable one $\bar{x}$.
Bottom: Basin of attraction to $\bar{x}$ in $x_1-x_2$ space (marked with ``X'') with filled
blue circles for convergent initial data and hollow black circles for divergent ones.}\label{fig:equilibria}
\end{figure}

\noindent
Class B networks are not expected to have a unique equilibrium. 
For instance the network of Figure \ref{fig:double_example} bottom with $ F_{i,h}(x_i)=\frac{x_i}{1+x_i}$ and $K_{i,h}(x_i)= \frac{1}{1+x_i}$ 
has two possible equilibria with values for $x_2$ given by
$\frac{f_3\, x_1}{f_1\,(1+x_1)}-1$
and $ \frac{f_2\,(1+x_1)}{f_4-f_2\,(1+x_1)}$. 
Choosing $f_{\{i=1,\ldots,7\}}= \{1,1,12,2,2,2,2\}$
the two equilibria have positive entries: $\tilde{x} = \{\frac{1}{2},3,1,1,1\}$ and $\bar{x} = \{\frac{1}{3},2,1,1,1\}$, see Fig \ref{fig:equilibria}.

%% file: Maxflow_example_graph.tex
\begin{tikzpicture}[every node/.style={scale=1},->,shorten >=1pt,auto,node distance=1cm,
        thick,main node/.style={circle,fill=blue!10,draw,minimum size=0.5cm,inner sep=0pt]}]
\node[main node] at (0,0) (0) {s};
\node[main node,right of=0] (1) {1};
\node[main node,right of=1] (2) {2};
\node[main node,above of=2] (3) {3};
\node[main node,above of=1] (4) {4};
\node[main node,right of=3,yshift=-0.5cm] (5) {t};



\path[-]
(0) edge[->,ultra thick] node[xshift=-0cm,yshift=0cm] {10} (1) ;
\path[-]
(1) edge[->,ultra thick] node[xshift=-0cm,yshift=-0.5cm] {$\infty$} (2) ;
\path[-]
(2) edge[->,ultra thick] node[xshift=0.5cm,yshift=-0.1cm] {$\infty$} (3) ;
\path[-]
(3) edge[->,ultra thick] node[xshift=-0cm,yshift=0cm] {$\infty$} (4) ;
\path[-]
(4) edge[->,ultra thick] node[xshift=-0cm,yshift=-0.1cm] {$\infty$} (1) ;
\path[-]
(3) edge[->,ultra thick] node[xshift=-0cm,yshift=0cm] {10} (5) ;

\end{tikzpicture} \hspace{1cm}  
\begin{tikzpicture}[every node/.style={scale=1},->,shorten >=1pt,auto,node distance=1cm,
        thick,main node/.style={circle,fill=blue!10,draw,minimum size=0.5cm,inner sep=0pt]}]
\node[main node] at (0,0) (0) {s};
\node[main node,right of=0] (1) {1};
\node[main node,right of=1] (2) {2};
\node[main node,above of=2] (3) {3};
\node[main node,above of=1] (4) {4};
\node[main node,right of=3,yshift=-0.5cm] (5) {t};

\node[draw=none] at (2.5,0.2) {$20$};

\path[-]
(0) edge[->,ultra thick] node[xshift=-0cm,yshift=0cm] {10} (1) ;
\path[-]
(1) edge[->,ultra thick] node[xshift=-0cm,yshift=-0.5cm] {$\infty$} (2) ;
\path[-]
(2) edge[bend right,->, ultra thick] node[xshift=-0cm,yshift=0cm] {} (3) ;
\path[-]
(2) edge[bend left,->, ultra thick] node[xshift=-0cm,yshift=0cm] {} (5) ;
\path[-]
(3) edge[->,ultra thick] node[xshift=-0cm,yshift=0cm] {$\infty$} (4) ;
\path[-]
(4) edge[->,ultra thick] node[xshift=-0cm,yshift=-0.1cm] {$\infty$} (1) ;

\end{tikzpicture}

%% file: Counter_example_graph.tex
%
%
%

\begin{tikzpicture}[every node/.style={scale=1.5},->,shorten >=1pt,auto,node distance=1cm,
        thick,main node/.style={circle,fill=blue!10,draw,minimum size=0.5cm,inner sep=0pt]}]
\node[main node] at (0,0) (1) {1};
\node[main node,right of=1] (2) {2};
\node[main node,above of=2] (3) {3};



\path[-]
(1) edge[->,ultra thick] (2) ;
\path[-]
(2) edge[->,ultra thick] (1) ;
\path[-]
(2) edge[->,ultra thick] (3) ;
\path[-]
(3) edge[bend right,->, ultra thick] node[xshift=-0cm,yshift = -0cm] {} (2) ;

\draw[|-,xscale=-1,ultra thick] (-1.8,.66) arc (90:280:.45);
\end{tikzpicture}


%% file: double_example.tex
\begin{tikzpicture}[every node/.style={scale=1},->,shorten >=1pt,auto,node distance=1cm,
        thick,main node/.style={circle,fill=blue!20,draw,minimum size=0.5cm,inner sep=0pt]}]
\node[draw=none] at (0,0) (source1) {};
\node[draw=none] at (0,1) (source2) {};
\node[main node] at (1,0) (1) {1};
\node[main node] at (1,1) (2) {2};
\node[main node] at (2.5,0) (3) {3};
\node[main node] at (2.5,1) (4) {4};
\node[main node] at (4,0) (5) {5};
\node[draw=none] at (5,0) (sink1) {};
\node[draw=none] at (4,1) (sink2) {};
\node[draw=none] at (2,2) (lab) {Class A};

	\path[-]
    (1) edge[->, ultra thick,below] node {$f_3$} (3)
    (3) edge[->, ultra thick,below] node {$f_5$} (5)
    (2) edge[->, ultra thick] node {$f_4$} (4)
    (source1) edge[->, ultra thick,below] node {$f_1$} (1)
    (source2) edge[->, ultra thick] node {$f_2$} (2)
    (5) edge[->, ultra thick,below] node {$f_7$} (sink1)
    (4) edge[->, ultra thick] node {$f_6$} (sink2);
    
    \draw[->||] (1) -- (1.75,0.9);
    \draw[-o] (4) -- (3.25,0);

\end{tikzpicture} \hspace{2cm}
\begin{tikzpicture}[every node/.style={scale=1},->,shorten >=1pt,auto,node distance=1cm,
        thick,main node/.style={circle,fill=blue!20,draw,minimum size=0.5cm,inner sep=0pt]}]
\node[draw=none] at (0,0) (source1) {};
\node[draw=none] at (0,1) (source2) {};
\node[main node] at (1,0) (1) {1};
\node[main node] at (1,1) (2) {2};
\node[main node] at (2.5,0) (3) {3};
\node[main node] at (2.5,1) (4) {4};
\node[main node] at (4,0) (5) {5};
\node[draw=none] at (5,0) (sink1) {};
\node[draw=none] at (4,1) (sink2) {};
\node[draw=none] at (2,2) (lab) {Class B};

	\path[-]
    (1) edge[->, ultra thick,below] node {$f_3$} (3)
    (3) edge[->, ultra thick,below] node {$f_5$} (5)
    (2) edge[->, ultra thick] node {$f_4$} (4)
    (source1) edge[->, ultra thick,below] node {$f_1$} (1)
    (source2) edge[->, ultra thick] node {$f_2$} (2)
    (5) edge[->, ultra thick,below] node {$f_7$} (sink1)
    (4) edge[->, ultra thick] node {$f_6$} (sink2);
    
    \draw[->||] (1) -- (1.75,0.9);
    \draw[->||] (2) -- (1.75,0.1);

\end{tikzpicture}

%% file: antibiotics_subgraph.tex
\begin{tikzpicture}[every node/.style={scale=0.9},->,shorten >=1pt,auto,node distance=1.0cm,
        thick,main node/.style={circle,fill=blue!20,draw,minimum size=0.5cm,inner sep=0pt]},scale=0.87]
\foreach \a in {1,2,...,10}{
\draw (\a*360/10-360/10: 3cm) node[main node](\a){\a};
}

    \node[draw=none] at (5.2, 2.8)  (a)     {Exchange fluxes};
    \node[draw=none] at (-4.5, -3)  (aa)     {Exchange fluxes};
    \node[draw=none] at (5, 1.6)  (a1)     {};
    \node[draw=none] at (0, 4.2)  (a2)     {};
    \node[draw=none] at (4.5, 1.8)  (a3)     {};
    \node[draw=none] at (0.4, 3.8)  (a4)     {};
    \node[draw=none] at (0.8, 1.8)  (b)     {\footnotesize Clofazimine};
    
	\node[draw=none] at (-1, 1.8)  (c)     {\footnotesize Amikacin,};
	\node[draw=none] at (-1, 1.4)  (d)     {\footnotesize Capreomycin};
	
	\node[draw=none] at (-1.4, -0.5)  (e)     {\footnotesize Rifapentine};
	
	\node[draw=none] at (-1, -1.6)  (f)     {\footnotesize Capreomycin};
	
	\node[draw=none] at (0.0, -2.2)  (g)     {\footnotesize Capreomycin};
	
	\node[draw=none] at (3.3, -2.2)  (h)     {\footnotesize Amikacin$^+$,};
	\node[draw=none] at (3.3, -2.6)  (i)     {\footnotesize Capreomycin$^+$,};
	\node[draw=none] at (3.1, -3.0)  (j)     {\footnotesize Clofazimine$^-$};
	
	\node[draw=none] at (1.6, -0.2)  (k)     {\footnotesize Amikacin$^+$,};
	\node[draw=none] at (1.4, -0.6)  (l)     {\footnotesize Capreomycin$^+$,};
	\node[draw=none] at (1.4, -1)  (m)     {\footnotesize Clofazimine$^-$};
	
	\node[draw=none] at (2.6, 3.7)  (n)     {\footnotesize Amikacin,};
	\node[draw=none] at (2.6, 3.4)  (o)     {\footnotesize Clofazimine,};
	\node[draw=none] at (2.6, 3.1)  (p)     {\footnotesize ETH, INH};

	\path[-]
    (1) edge[->,bend right, ultra thick] node {} (2)
    (2) edge[->,bend right, ultra thick] node {} (3)
    (3) edge[->,bend right, ultra thick] node {} (4)
    (4) edge[->,bend right, ultra thick] node {} (5)
    (5) edge[->,bend right, ultra thick] node {} (6)
    (6) edge[->,bend right, ultra thick] node {} (7)
    (7) edge[->,bend right, ultra thick] node {} (8)
    (8) edge[->,bend right, ultra thick] node {} (9)
    (9) edge[->,bend right, dotted, line width=3pt, blue] node {} (10)
    (1) edge[->,bend right, dotted, line width=3pt, blue] node {} (10)
    (2) edge[->,bend right, ultra thick] node {} (1)
    (3) edge[->,bend right, ultra thick,dashed, red] node {} (2)
    (4) edge[->,bend right, ultra thick] node {} (3)
    (5) edge[->,bend right, ultra thick,dashed, red] node {} (4)
    (6) edge[->,bend right, ultra thick] node {} (5)
    (7) edge[->,bend right, ultra thick,dashed, red] node {} (6)
    (8) edge[->,bend right, ultra thick,dashed, red] node {} (7)
    (9) edge[->,bend right, ultra thick,dashed, red] node {} (8)
    (10) edge[->,bend right, ultra thick] node {} (9)
    (10) edge[->,bend right, ultra thick] node {} (1)
    (2) edge[->, bend left,  line width=4pt] node {} (a)
    (a) edge[->, bend left,ultra thick] node {} (2)
    (a1) edge[->, bend left,ultra thick] node {} (1)
    (1) edge[->, bend right,ultra thick] node {} (a3)
    (a2) edge[->, bend right,ultra thick] node {} (4)
    (4) edge[->, bend left,ultra thick] node {} (a4)
    (7) edge[->,bend right,ultra thick] node {} (aa)
    (aa) edge[->,bend right,ultra thick] node {} (7);

\end{tikzpicture}